\documentclass[
twocolumn,
aps,
prl,
reprint,
superscriptaddress,
amsmath,amssymb,
nofootinbib
]{revtex4-2}
\renewcommand{\selectlanguage}[1]{}

\usepackage[pdftex]{graphicx,color}
\usepackage[pdftex,
            colorlinks=true,
            citecolor=blue,
            linkcolor=blue]{hyperref}

\graphicspath{{./figure},{./fig_SM}}

\usepackage{braket}
\usepackage{times,multirow,amsfonts,bm,xspace,pifont}
\usepackage{soul}
\usepackage[normalem]{ulem}

\begin{document}
\newcommand{\sgn}{\mathrm{sgn}\,}
\newcommand{\tr}{\mathrm{tr}\,}
\newcommand{\Tr}{\mathrm{Tr}\,}

\newcommand{\YY}[1]{\textcolor{magenta}{#1}}
\newcommand{\YYY}[1]{\textcolor{green}{#1}}
\newcommand{\AD}[1]{\textcolor{blue}{#1}}
\newcommand*{\ADS}[1]{\textcolor{blue}{\sout{#1}}}
\newcommand*\YYS[1]{\textcolor{magenta}{\sout{#1}}}
\newcommand{\reply}[1]{{#1}}
\newcommand{\replyS}[1]{\textcolor{red}{\sout{#1}}}

\renewcommand{\Re}{\mathrm{Re}}
\renewcommand{\Im}{\mathrm{Im}}
\newcommand{\diag}{\mathrm{diag}}

\renewcommand{\jcp}{{j_{\rm c+}}}
\newcommand{\jcm}{{j_{\rm c-}}}
\newcommand{\jeq}{{j_{\rm eq}}}
\newcommand{\jss}{{j_{\rm ss}}}
\newcommand{\mE}{{\mathcal{E}}}
\newcommand{\mEc}{{\mathcal{E}_{\rm c}}}
\newcommand{\mEct}{{\mathcal{E}_{\rm c2}}}
\newcommand{\mC}{{\mathcal{C}}}
\newcommand{\sst}{{\rm ss}}
\newcommand{\mP}{\mathcal{P}}
\newcommand{\mQ}{\mathcal{Q}}
\newcommand{\mM}{\mathcal{M}}
\newcommand{\bk}{\bar{k}}

\newcommand{\Tc}{{T_{\rm c}}}
\newcommand{\qc}{{q_{\rm c}}}
\newcommand{\Tcn}{{T_{\rm c0}}}

\newcommand{\bkpsit}[1]{\braket{#1}_{\hat{\psi}(q,t)}}
\newcommand{\bkpsi}[1]{\braket{#1}_{\hat{\psi}}}
\newcommand{\alphaeff}{\hat{\alpha}_{\rm eff}}
\newcommand{\halpha}{\hat{\alpha}}
\newcommand{\hgamma}{\hat{\gamma}}
\newcommand{\hbeta}{\hat{\beta}}
\newcommand{\rbra}[1]{{(#1|}}
\newcommand{\rket}[1]{{|#1)}}
\newcommand{\rbraket}[1]{{(#1\rangle}}

\title{Unidirectional superconductivity and {superconducting} diode effect induced by dissipation}

\author{Akito Daido}
\affiliation{Department of Physics, Graduate School of Science, Kyoto University, Kyoto 606-8502, Japan}
\email[]{daido@scphys.kyoto-u.ac.jp}
\author{Youichi Yanase}
\affiliation{Department of Physics, Graduate School of Science, Kyoto University, Kyoto 606-8502, Japan}
\date{\today}

\begin{abstract}
A general principle of condensed matter physics prohibits the electric current in equilibrium. 
This prevents a zero-resistance state realized solely under a finite electric current, namely unidirectional superconductivity.
In this {paper}, we propose a setup to realize the unidirectional superconductivity as a nonequilibrium steady state.
We focus on the in-plane transport of atomically thin bilayer superconductors lacking in-plane inversion symmetry and introduce dissipation by applying the out-of-plane electric field and current.
By analyzing time-dependent Ginzburg-Landau equations, we show that locally stable steady-state solutions appear only under an in-plane supercurrent when the out-of-plane electric field exceeds a threshold value.
Our system also realizes a dissipation-induced superconducting diode effect up to $100\%$ efficiency by purely electric means.
\end{abstract}

\maketitle

\begin{figure}
    \centering
    \includegraphics[width=0.4\textwidth]{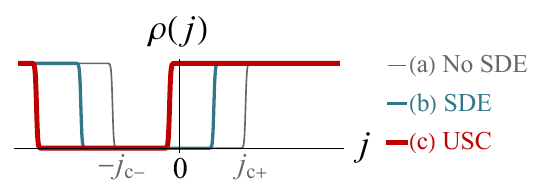}
    \caption{
    Schematic of current-resistivity relation $\rho(j)$ 
{in three cases: (a) no SDE, (b) SDE, and (c) USC. The USC can be induced by dissipation and realizes a perfect SDE.}
    }
    \label{fig:USC}
\end{figure}

\textit{Introduction}. --- 
Current-related phenomena in superconductors are among the most important topics in condensed matter physics.
A recent experiment of the superconducting diode effect (SDE) in Nb/V/Ta superlattices~\cite{Ando2020-om} has brought about a number of theoretical and experimental studies with particular emphasis on the potential role of bulk inversion symmetry breaking~\cite{Wu2022-ey,Bauriedl2022-nq,Narita2022-od,Yun2023-je,Lin2022-cz,Scammell2022-pv,Banerjee2024-kd,Yuan2022-pz,Daido2022-ox,He2022-px,Ilic2022-kh,Daido2022-gj}.
SDE refers to the situation
where the critical currents for the positive and negative directions, namely $j_{\rm c\pm}$, are different from each other [Fig.~\ref{fig:USC}(b)].
It has been pointed out that not only vortex-related extrinsic mechanisms~\cite{Jiang1994-yy,Ichikawa1994-pl,Vodolazov2005-cg,Harrington2009-zj,
Lyu2021-sm,Hou2023-hu,Mizuno2022-bd} but also an intrinsic mechanism by finite-momentum Cooper pairs can cause SDE~\cite{Yuan2022-pz,Daido2022-ox,He2022-px,Ilic2022-kh,Daido2022-gj,Scammell2022-pv,Banerjee2024-kd}, shedding light on the potential application of SDE for probing exotic superconducting states.
Nonreciprocal resistivity during the superconducting transition is also attracting much attention~\cite{Tokura2018-nb,Ideue2021-es,Wakatsuki2017-dp,Wakatsuki2018-ll,Hoshino2018-sa,
Qin2017-vd,Yasuda2019-jw,Zhang2020-al,Daido2024-pw}.
Further study of the current-related phenomena in superconductors is indispensable both from fundamental-physics and engineering viewpoints.

Among various experiments of SDE, a notable result has been reported for twisted trilayer graphene~\cite{Lin2022-cz,Scammell2022-pv,Banerjee2024-kd}. In this system,
either $j_{\rm c+}$ or $j_{\rm c-}$ vanishes as in Fig.~\ref{fig:USC}(c) for some values of the carrier density.
The result is surprising because the electric current stabilizes superconductivity rather than breaking it.
Such a situation does not usually occur because a current-free state is ensured to be the most stable~\cite{Bohm1949-gk,Ohashi1996-vu,Watanabe2019-qh}.
Specifically, in the mean-field approximation, the free energy $F$, supercurrent $j$, and Cooper-pair momentum $q$ are related to each other by $j=2\partial F/\partial q$~\cite{Tinkham2004-dh}.
This ensures $j=0$ to be the global minimum of the free energy and also ensures the presence of superconducting states with $j>0$ and $j<0$ nearby.
Thus, $j_{\rm c\pm}$ can not vanish, although it could be tiny.
This conclusion remains unchanged even with vortices, whose depinning requires a finite electric current.

We must circumvent the above discussion to achieve a zero-resistance state only under a finite electric current, namely, unidirectional superconductivity (USC).
One way is to assume 
a strong dependence of system parameters on the electric current.
Actually, it has been pointed out that USC in twisted trilayer graphene might be understood by destabilization of the coexisting symmetry-breaking order detrimental to superconductivity~\cite{Scammell2022-pv}.
\begin{figure}
    \centering
    \includegraphics[width=0.45\textwidth]{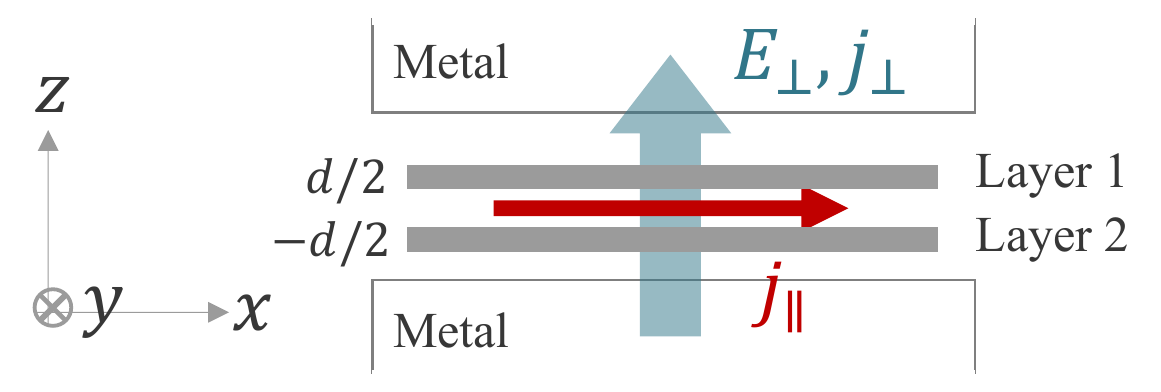}
    \caption{Schematic figure of the system.
    We consider a wire or thin film uniform in the $y$ direction made of a bilayer superconductor and apply the electric field $E_\perp$ (accompanying electric current $j_\perp$) in the $z$ direction. We discuss the supercurrent $j_\parallel$ in the $x$ direction.
    }
    \label{fig:system}
\end{figure}
Such a scenario, however, relies on nontrivial assumptions of the underlying exotic order and is difficult to quantify.
This stands in the way of predicting new USC candidates {and discovering high-performance superconducting diodes}.

In this {paper}, we propose another direction to achieve USC, i.e., driving the system out of equilibrium.
As a source of dissipation, we focus on the dc electric field and current, which should be applied perpendicular to the system to study the in-plane supercurrent [Fig.~\ref{fig:system}].
Generally speaking, dc electric field penetrates into superconductors on a length scale comparable to {or larger than} the coherence length~\cite{Kopnin2001-ie,Artemenko1979-js,Ivlev1984-od}, while it vanishes in the bulk.
Thus, 
superconducting wires and films 
{with a thickness comparable to the electric-field penetration depth, sandwiched with normal-metal leads [Fig.~\ref{fig:system}],} would provide the platform for USC.
{Such a superconductor junction has long been studied as a central platform of nonequilibrium superconductivity~\cite{Kopnin2001-ie,
Artemenko1979-js,Ivlev1984-od,Stoof1996-uq,Vodolazov2003-nv,Boogaard2004-ro,Keizer2006-oh,Rubinstein2007-ck,Snyman2009-jg,Moor2009-mx,Catelani2010-yv,Vercruyssen2012-uw,Chtchelkatchev2012-jd,
Bobkova2013-le,Serbyn2013-ji,Ouassou2018-mj,Seja2021-na,Kawamura2024-ot}.
}
{It should be noted that this setup is slightly different from the electrostatic gating, where the perpendicular electric field, if any, creates only a tiny leakage current due to the insulating layers.
}

In the following, we focus on atomically thin bilayer superconductors as a minimal {toy model, while}
{we expect qualitatively the same results for thicker films}.
From the symmetry viewpoint, 
in-plane inversion and time-reversal symmetries
must be broken to achieve SDE~\cite{Yuan2022-pz,Daido2022-ox,He2022-px,Ilic2022-kh,Daido2022-gj} and therefore USC.
We require the crystal structure to satisfy the former
while the latter is accomplished by dissipation.
We demonstrate that such a bilayer superconductor actually realizes {dissipation-induced SDE and} USC as a nonequilibrium steady state, based on the time-dependent Ginzburg-Landau (TDGL) equation.
Our result uncovers a scheme to realize ideal {field-free} superconducting diodes. 

\textit{Bilayer superconductor in equilibrium.} --- 
Let us first discuss the equilibrium properties in the absence of the perpendicular electric field [Fig.~\ref{fig:system}].
We consider a one-dimensional superconducting channel of a bilayer superconductor described by the Ginzburg-Landau (GL) free energy
\begin{equation}
  \!\!\!\!  F[\bm{\psi}]=N_0\!\!\int\! dx\,\bm{\psi}^\dagger(x)\hat{\alpha}(\nabla/i)\bm{\psi}(x)+\frac{\beta_0}{2}[\bm{\psi}^\dagger(x)\bm{\psi}(x)]^2,\label{eq:GL}
\end{equation}
where each component of $\bm{\psi}(x)=\bigl(\psi_1(x),\psi_2(x)\bigr)^T$
corresponds to the order parameter of layers 1 and 2 at $z=\pm d/2$.
$N_0$ and $\beta_0$ are positive constants.
The first term includes the matrix
$\hat{\alpha}(\nabla/i)=\alpha_0(\nabla/i)+\bm{\alpha}(\nabla/i)\cdot\bm{\sigma}$,
where $\sigma_i$ denotes Pauli matrices in the layer space.

It is convenient to introduce the plane-wave basis, replacing
${\nabla}/i$ with the center-of-mass momentum $q$ of Cooper pairs.
By assuming the time-reversal symmetry $\hat{\alpha}^*(q)=\hat{\alpha}(-q)$, the components $\alpha_0(q)$, $\alpha_x(q)$, and $\alpha_z(q)$ are even functions of $q$, while $\alpha_y(q)$ is an odd function.
As a minimal model, we consider
$\alpha_0(q)=a_0+a_0''q^2,
\alpha_x(q)=a_x,
\alpha_y(q)=a_y'q,$
and the high-symmetry quartic term in Eq.~\eqref{eq:GL} {(see Supplemental Material~\cite{Supplemental} for a generalization)}.
Here, $\alpha_z(q)=0$ by imposing the global inversion symmetry $\sigma_x\hat{\alpha}(q)\sigma_x=\hat{\alpha}(-q)$, which is to be broken under the perpendicular electric field.
The term $a_y'=\partial_q\alpha_y(q)|_{q\to0}$ requires the absence of the $x$-mirror plane and out-of-plane rotational axis [i.e., the in-plane inversion symmetry].
Thus, the point-group symmetry of our model is maximally $C_{2h}$ with the principal axis along the $y$ direction,
while its subgroups 
can also show USC~\cite{Supplemental}.
We also introduce $\alpha_\rho(q)$ and $\alpha_\theta(q)$ by {$(\alpha_x(q), \alpha_y(q)) = \alpha_\rho(q)(\cos\alpha_\theta(q), \sin\alpha_\theta(q))$ for simplicity of notation.}

The system undergoes a superconducting transition when the minimum eigenvalue of $\hat{\alpha}(\nabla/i)$ becomes negative as lowering the temperature $T$.
For a given $q$, $\hat{\alpha}(q)$ has two branches $\alpha_\pm(q)=\alpha_0(q)\pm\alpha_\rho(q)$, and the lower branch is expanded as
$\alpha_-(q)\simeq \epsilon+\xi_0^2q^2+O(q^4)$.
This is minimized at $q=0$,
assuming the squared coherence length $\xi_0^2\equiv a_0''-a_y'^2/2|a_x|$ to be positive. The transition temperature ${T_{\rm c0}}$ is reached when the reduced temperature $\epsilon\equiv (T-{T_{\rm c0}})/{T_{\rm c0}}=a_0-|a_x|$ vanishes.
The normalized eigenstate of the lower branch is 
\begin{equation}
     \hat{\psi}_{\rm eq}(q)\simeq\frac{1}{2}e^{-\frac{i}{2}\left[\alpha_\theta(0)+\frac{a_y'}{a_x}q\right]\sigma_z}
\begin{pmatrix}
    1\\-1
\end{pmatrix}.
    \label{eq:equilibriumstates}
\end{equation}
The Bardeen-Cooper-Schrieffer state $\hat{\psi}_{\rm eq}({0})\propto (1,1)^T$ or  pair-density-wave state~\cite{Khim2021-er,Fischer2023-rf} $\hat{\psi}_{\rm eq}(0)\propto (1,-1)^T$ is stabilized depending on the sign of $a_x$, which does not affect the following discussion. 
Importantly, the relative phase between layers
is modified for 
$q\neq0$.
It will be shown that this effect combined with the electric field gives a significantly nonreciprocal current response and results in USC.

The amplitude of the superconducting order parameter becomes finite for a negative reduced temperature $\epsilon<0$.
The full superconducting solution $\bm{\psi}({x})$ can be obtained by solving the GL equation
\begin{equation}
    \frac{\delta F}{\delta\bm{\psi}^\dagger(x)}=
    N_0\Bigl\{\hat{\alpha}(\nabla/i)+\beta_0[\bm{\psi}^\dagger(x)\bm{\psi}(x)]\Bigr\}\bm{\psi}(x)=0.\label{eq:GLequation}
\end{equation}
This can be easily solved for the plane-wave ansatz.
By assuming $\bm{\psi}({x})=e^{iqx}R(q)\hat{\psi}(q)$ with $R(q)\ge0$ and $|\hat{\psi}(q)|=1$, $\hat{\psi}(q)$ turns out to be a normalized eigenstate of $\hat{\alpha}(q)$.
Choosing 
the branch $\hat{\psi}_{\rm eq}(q)$, the order-parameter amplitude satisfies $R_{\rm eq}(q)^2=(-\epsilon-\xi_0^2q^2)/\beta_0$, which must be positive for the state with momentum $q$ to exist.

The supercurrent through the system is obtained by~\cite{Supplemental}
\begin{equation}
j_\parallel(q)
=2N_0R^2(q)\hat{\psi}^\dagger(q)\partial_{q}\hat{\alpha}(q)\hat{\psi}(q),
\label{eq:current_q}
\end{equation}
for any plane-wave order parameter.
We obtain
\begin{equation}
    j_{\rm eq}(q)\simeq\frac{4N_0\xi_0^2}{\beta_0}\,q\,(|\epsilon|-\xi_0^2q^2),
\end{equation}
for the equilibrium solution.
This is the conventional current-momentum relation~\cite{Tinkham2004-dh}.
The maximum ${j_{\rm c+}}$ and minimum $-{j_{\rm c-}}$ of ${j_{\rm eq}}(q)$ give the depairing critical currents for positive and negative directions, exceeding which no superconducting solution can support the applied electric current.
The equilibrium critical current ${j_{\rm c+}}={j_{\rm c-}}$ is reciprocal 
in accordance with the global inversion and time-reversal symmetries, both of which are absent under the perpendicular electric field as clarified below. 

\textit{Nonequilibrium steady states.} --- 
Next, we discuss the nonequilibrium properties of the system.
The nonequilibrium dynamics of the order parameter can be phenomenologically understood by the TDGL equation
\begin{equation}
    \Gamma\partial_t\bm{\psi}(x,t)=-\left.\frac{\delta F}{\delta\bm{\psi}^\dagger(x,t)}\right.,\label{eq:TDGL}
\end{equation}
where $\Gamma>0$ and $\partial_t\equiv\partial/\partial t$.
The TDGL equation describes a relaxation dynamics of the order parameter toward a steady state, that is, the solution of the GL equation~\eqref{eq:GLequation}.
While the TDGL equation is microscopically justified only in limited situations including gapless superconductors~\cite{Kopnin2001-ie,Artemenko1979-js,
Ivlev1984-od},
the simplified description offered by the TDGL formalism is useful to qualitatively understand and predict physical phenomena.

The effect of the electromagnetic field can be incorporated into Eq.~\eqref{eq:TDGL} by 
considering the gauge invariance of the equation~\cite{Supplemental}.
In particular, the electric field can be introduced by $\partial_t\to \partial_t+2i\phi$ with the scalar potential $\phi$.
While the in-plane dc electric field $E_\parallel$ along the system can be considered by $\phi=-E_\parallel x$ as in Refs.~\cite{Rubinstein2007-ck, Chtchelkatchev2012-jd,
Serbyn2013-ji,
Artemenko1979-js,Kopnin2001-ie}, we focus on the dc electric field perpendicular to the system by $\phi=-E_\perp d\sigma_z/2$:
Since the thickness $d$ is 
{a microscopic}-scale distance, $E_\perp$ can exist in the system and accompany the perpendicular flow of Cooper pairs [Fig.~\ref{fig:system}] and normal electrons
~\cite{Kopnin2001-ie}.
Thus, assuming the plane-wave ansatz $\bm{\psi}(x,t)=e^{iqx}\bm{\psi}(q,t)$, the TDGL equation recasts into~\cite{Supplemental}
\begin{equation}
[\partial_t+i\Phi]\bm{\psi}(q,t)=-\left[\hat{\alpha}(q)+\beta_0|\bm{\psi}(q,t)|^2 \right]\bm{\psi}(q,t),\label{eq:TDGL_bl}
\end{equation}
with the dimensionless scalar potential $\Phi=-{\mathcal{E}}\sigma_z$ and electric field ${\mathcal{E}}\equiv\Gamma E_\perp d/N_0$.
Here and hereafter, the time $t$ is rescaled by $\Gamma/N_0$.

Let us search for the steady-state solution of Eq.~\eqref{eq:TDGL_bl}.
By writing $\bm{\psi}(q,t)=e^{i\chi(q,t)}R(q)\hat{\psi}(q)$,
we require that only the global phase factor $\chi(q,t)$, which is irrelevant for physical quantities, may be dependent on time~\cite{Rubinstein2007-ck}.
Then, the nonequilibrium steady state is given by the right eigenstate of a non-Hermitian matrix $\hat{\alpha}_{\rm eff}(q)=\halpha(q)+i\Phi$,
whose eigenvalues are $\lambda_\pm(q)=\alpha_0(q)\pm\sqrt{\alpha_\rho(q)^2-{\mathcal{E}}^2}$.
The spectrum remains real {for superconducting solutions} in our model owing to the parity-time-reversal symmetry {$\sigma_x\hat{\alpha}_{\rm eff}(q)^*\sigma_x=\hat{\alpha}_{\rm eff}(q)$.}
After diagonalizing $\alphaeff(q)$~\cite{Supplemental}, the nonequilibrium steady state is explicitly written as $\bm{\psi}(q,t)=e^{i\chi_{\rm ss}(q,t)}R_{\rm ss}(q)\hat{\psi}_{\rm ss}(q)$ and
\begin{subequations}\begin{gather}
R_{\rm ss}(q)^2\simeq\frac{1}{\beta_0}\left[-\bar{\epsilon}({\mathcal{E}})-\xi_0^2q^2\right],\label{eq:R_steadystate}\\
\hat{\psi}_{\rm ss}(q)\simeq\frac{1}{2}e^{-\frac{i}{2}\left[\alpha_\theta(0)+\frac{a_y'}{a_x}q-\theta_{\mathcal{E}}\right]\sigma_z}
\begin{pmatrix}
    1\\-1
\end{pmatrix},
\label{eq:psi_steadystate}
\end{gather}\label{eq:steadystate_expression}\end{subequations}
with $\bar{\epsilon}({\mathcal{E}})\equiv\epsilon+{{\mathcal{E}}^2}/{2|a_x|}$.
Compared with Eq.~\eqref{eq:equilibriumstates}, an extra phase difference $\theta_{\mathcal{E}}\simeq {\mathcal{E}}/|a_x|$ appears in $\hat{\psi}_{\rm ss}(q)$.
The global phase factor of the steady state turns out to be time-independent,
$\partial_t\chi_{\rm ss}(q,t)=-\mathrm{Im}[\lambda_-(q)]=0$,
{by the} parity-time-reversal symmetry.

\textit{Supercurrent carried by the steady states.} --- 
So far, we have derived the steady states of the TDGL equations, which are modified from the equilibrium state Eq.~\eqref{eq:equilibriumstates} by the phase factor $\theta_{\mathcal{E}}$.
Its appearance in $\hat{\psi}_{\rm ss}(q)$ can be understood 
by considering the perpendicular Cooper-pair current driven by the electric field,
$j_\perp(q)\simeq 2N_0R_{\rm ss}(q)^2|a_x|\sin\theta_{\mathcal{E}}$~\cite{Supplemental}.
The presence of $j_\perp(q)$ naturally ensures the inter-layer phase difference $\theta_{\mathcal{E}}$ in analogy with the Josephson effect.
In the following, we clarify that the phase $\theta_{\mathcal{E}}$ is the key to realizing USC, in combination with the GL coefficient $a_y'=\partial_q\alpha_y(q)|_{q\to0}$.

Before discussing the current-related properties, we show the temperature scaling of the quantities.
The electric field is detrimental to superconductivity
since ${\mathcal{E}}$ effectively increases the reduced temperature by $\epsilon\to\bar{\epsilon}({\mathcal{E}}) = \epsilon+{{\mathcal{E}}^2}/{2|a_x|}$. This causes a transition into the normal state when $\bar{\epsilon}({\mathcal{E}_{\rm c}})=0$ with the critical electric field 
\begin{equation}
{\mathcal{E}}_{\rm c}(\epsilon)\equiv \sqrt{2|a_x|\,|\epsilon|}{=\mathcal{E}_{\rm c}(-1)\sqrt{|\epsilon|}}.
\end{equation}
Accordingly, the critical voltage 
$\sim 8T_{\rm c0}\sqrt{2|a_x|\,|\epsilon|}/\pi$
is of the order $T_{\rm c0}$
by using $\Gamma/N_0\sim\pi/8T_{\rm c0}$~\cite{Larkin2005-lb}.
We are interested in the regime 
$|{\mathcal{E}}|\le{\mathcal{E}_{\rm c}}(\epsilon)$.
We also limit ourselves to the momentum regime $|\xi q|<1$ with $\xi\equiv\xi_0/\sqrt{|\bar{\epsilon}({\mathcal{E}})|}$,
outside which 
steady states do not exist as follows from Eq.~\eqref{eq:R_steadystate}.
The $|\epsilon|^{1/2}$ scaling of $\mE$ and $q$ ensures the relation
$|{\mathcal{E}}|$, $|a_y'q|\ll|a_x|$, which was used to derive Eqs.~\eqref{eq:steadystate_expression}.

\begin{figure}
    \centering
    \includegraphics[width=0.45\textwidth]{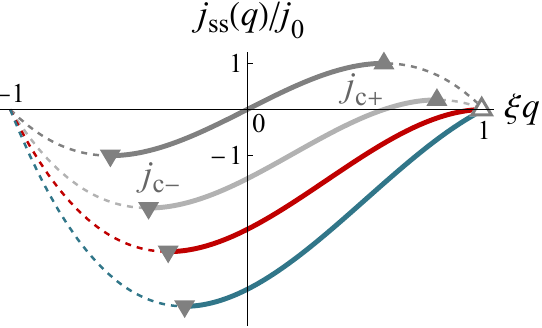}
    \caption{The current-momentum relation $j_{\rm ss}(q)$ of the nonequilibrium steady state for various values of the perpendicular electric field ${\mathcal{E}}$. 
    The vertical axis is ${j_{\rm ss}}(q)/j_{\rm 0}$ with $j_{\rm  0}=8N_0\xi_0|\bar{\epsilon}({\mathcal{E}})|^{3/2}/3\sqrt{3}\beta_0$, and the horizontal axis is $\xi q$.
    Solid and dashed lines show locally stable and unstable states, respectively.
    The dark and light gray, red, and blue lines correspond to $\xi q_{\mathcal{E}}=0,\frac{1}{\sqrt{3}}, 1,$ and $1.5$, respectively.
    Triangles indicate ${j_{\rm c+}}$, which vanishes for $\xi q_{\mathcal{E}}\ge1$ [open triangle], while reversed triangles indicate ${j_{\rm c-}}$.
    }
    \label{fig:jq_ss}
\end{figure}

The current-momentum relation of the nonequilibrium steady state can be obtained by plugging the order parameter $\bm{\psi}(q,t)=e^{i\chi_{\rm ss}(q,t)}R_{\rm ss}(q)\hat{\psi}_{\rm ss}(q)$ into Eq.~\eqref{eq:current_q}.
It should be noted that the nonequilibrium modification of the order parameter
is understood as a shift in momentum,
\begin{equation}
\hat{\psi}_{\rm ss}(q)\simeq \hat{\psi}_{\rm eq}(q-a_x\theta_{\mathcal{E}}/a_y'),
\end{equation}
while the amplitude $R_{\rm ss}(q)$ does not experience such a shift.
This means that the GL coefficient $a_y'=\partial_q\alpha_y(q)|_{q\to0}$ behaves as a translator between the electric field and the Cooper-pair momentum,
which naturally leads to a momentum shift in the current-momentum relation as well.
After partial cancellation with the other contribution~\cite{Supplemental}, 
the current-momentum relation of the steady state is given by,
\begin{equation}
{j_{\rm ss}}(q)=\frac{4N_0\xi_0^2}{\beta_0}(q-q_{\mathcal{E}})(|\bar{\epsilon}({\mathcal{E}})|-\xi_0^2q^2),\label{eq:jss}
\end{equation}
with the pumped momentum $q_{\mathcal{E}}=-a_y'{\mathcal{E}}/{2\xi_0^2a_x}$.
From a technical viewpoint, the appearance of $q_{\mathcal{E}}$ is due to the non-orthogonality of non-Hermitian eigenstates~\cite{Supplemental,Ashida2020-yn}, and thus reflects the nonequilibrium nature of the steady state.

In Fig.~\ref{fig:jq_ss}, we show the in-plane supercurrent $j_{\rm ss}(q)$ for several values of the out-of-plane electric field ${\mathcal{E}}$.
The dark gray line indicates $j_{\rm eq}(q)$ in the absence of the electric field, which evolves downward by applying the electric field [light gray, red, and blue lines].
The solid (dashed) lines indicate the locally stable (unstable) steady states,
which guarantees that
any small perturbation
decays in time.
By analyzing the linearized TDGL equation under the current bias, the local-stability condition turns out to be $dj_\sst(q)/dq>0$~\cite{Supplemental}, which naturally generalizes the condition $dj_{\rm eq}(q)/dq>0$ in equilibrium~\cite{Langer1967-rh,Samokhin2017-xf}.
Interestingly, this does not coincide with the local-stability condition $|\xi q|<1/\sqrt{3}$ 
without the current bias~\cite{Supplemental}.

We find an interesting feature when focusing on the solid lines in Fig.~\ref{fig:jq_ss}: The maximum current carried by the locally stable steady states vanishes by increasing the electric field ${\mathcal{E}}$.
The red line corresponds to the system just under the threshold electric field ${\mathcal{E}_{\rm c2}}(\epsilon)$, where the critical current ${j_{\rm c+}}$ in the positive direction vanishes.
Afterward, the system can carry only the negative supercurrent, or equivalently, shows zero resistance only under the negative electric current.
This indicates the realization of USC.

The onset electric field ${\mathcal{E}_{\rm c2}}(\epsilon)$ of USC can be obtained by the condition $|\xi q_{\mathcal{E}}|=1$.
The ratio of ${\mathcal{E}_{\rm c2}}(\epsilon)$ to the critical electric field ${\mathcal{E}_{\rm c}}(\epsilon)$ is~\cite{Supplemental}
\begin{equation}
\frac{{\mathcal{E}_{\rm c2}}(\epsilon)}{{\mathcal{E}_{\rm c}}(\epsilon)}=\frac{1}{\sqrt{1+{\mathcal{C}}^2}}\le1,\label{eq:Ec2byC}
\end{equation}
where we defined
\begin{equation}
{\mathcal{C}}\equiv\xi_0\frac{dq_{\mathcal{E}}}{d{\mathcal{E}}}
{\mathcal{E}_{\rm c}(-1)}
={\pm\frac{|\xi_0^{-1}a_y'|}{\sqrt{2|a_x|}},}
\end{equation}
with $\pm=\mathrm{sgn}\,[-a_x a_y']$.
Equation~\eqref{eq:Ec2byC} indicates that USC always occurs before the superconductivity is completely destroyed at the critical electric field.

\begin{figure}  
\flushleft{(a)$\qquad\qquad\qquad\qquad\qquad$(b)}\\
\includegraphics[width=0.5\textwidth]{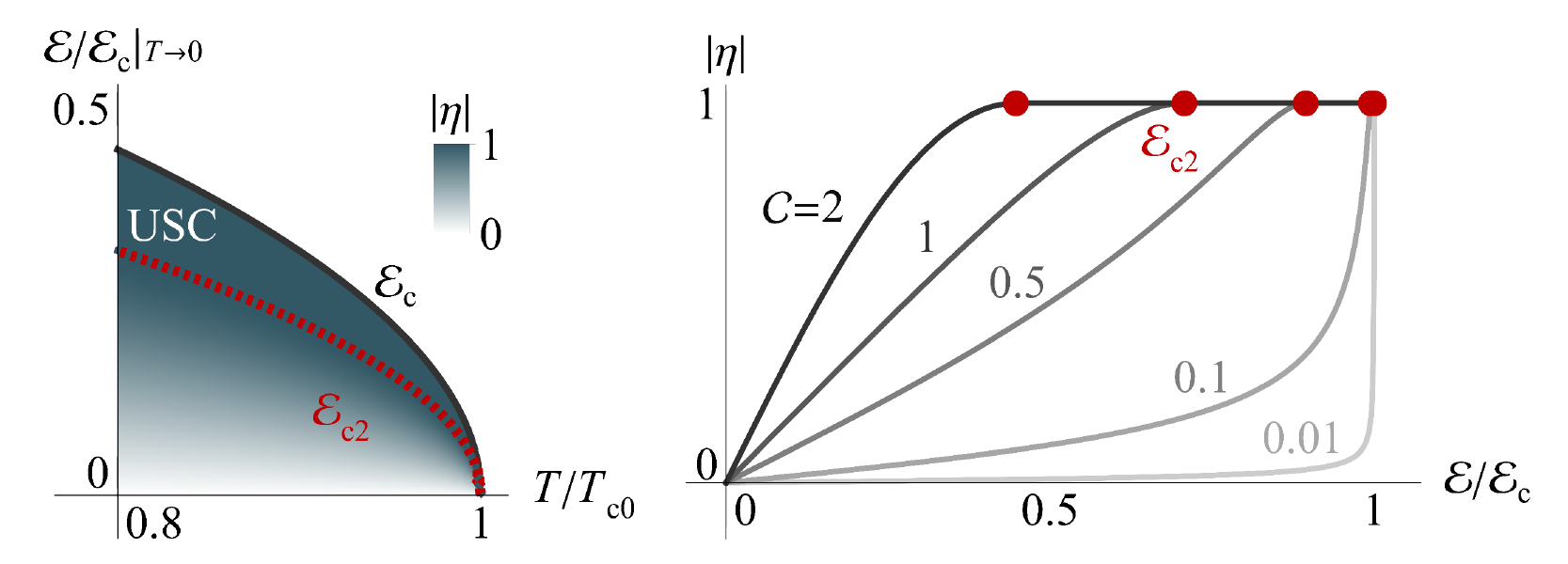}
    \caption{(a) The phase diagram of USC for ${\mathcal{C}}=1$.
    The horizontal and vertical axes are $T/T_{\rm c0}=1+\epsilon$ and ${\mathcal{E}}/\left.{\mathcal{E}}_{\rm c}\right|_{T\to0}$.
    The blue gradation indicates the diode quality factor $|\eta|$.
    (b) The diode quality factor $|\eta|$ at $T/T_{\rm c0}=0.9$ for various values of ${\mathcal{C}}=0.01,$ $0.1$, $0.5$, $1$, and $2$.
    The horizontal and vertical axes are ${\mathcal{E}}/\left.{\mathcal{E}}_{\rm c}\right|_{T=0.9T_{\rm c}}$ and $|\eta|$.
    The red points indicate the onset electric field for USC, ${\mathcal{E}}=\left.{\mathcal{E}}_{\rm c2}\right|_{T=0.9T_{\rm c}}$.
    }
    \label{fig:phasediagram}
\end{figure}

The dimensionless constant ${\mathcal{C}}$ describes the conversion efficiency between the electric field and Cooper-pair momentum and is the key parameter determining the nonreciprocity.
We show in Fig.~\ref{fig:phasediagram}(a) the phase diagram for ${\mathcal{C}}=1$ near the transition temperature $0.8\le T/{T_{\rm c0}}\le1$, in which the blue gradation indicates the absolute value of the diode quality factor $\eta\equiv({j_{\rm c+}}-{j_{\rm c-}})/({j_{\rm c+}}+{j_{\rm c-}})$ for the depairing critical current.
USC is realized for the electric field ${\mathcal{E}}_{\rm c2}(\epsilon)\le{\mathcal{E}}\le{\mathcal{E}}_{\rm c}(\epsilon)$, where $|\eta|=1$.
A sizable SDE is obtained even below ${\mathcal{E}_{\rm c2}}(\epsilon)$.
We also show in Fig.~\ref{fig:phasediagram}(b) the evolution of $\eta$ for various values of ${\mathcal{C}}$ when ${\mathcal{E}}$ is increased.
A large portion of the superconducting region 
is occupied by USC for ${\mathcal{C}}\gtrsim0.5$.
The quality factor $\eta$ for small electric field $\mathcal{E}$ is given by
\begin{align}
\eta&=-\sqrt{3}\,\mC\,\frac{\mE}{\mE_{\rm c}(\epsilon)}+O\left({\mE}^2\right),\label{eq:eta_low_E}
\end{align}
describing the dissipation-induced SDE.
This is
{much larger than the}
nonreciprocity of the equilibrium depairing critical current{, which decays as approaching the transition line}~\cite{Yuan2022-pz,Daido2022-ox,He2022-px,Ilic2022-kh,Daido2022-gj}.

\textit{Discussion.} --- Our results are straightforwardly generalized to arbitrary bilayer GL models~\cite{Supplemental}.
{In particular, the steady state is approximated by the eigenstate of $\hat{\alpha}_{\rm eff}(q)$ even with nontrivial GL quartic terms.}
Thus, Eq.~\eqref{eq:jss} and its subsequent discussion remain valid by reinterpreting $\bar{\epsilon}({\mathcal{E}})$, $\xi_0^2$, $\beta_0$, and $q_{\mathcal{E}}$ as those evaluated by the GL model under consideration, ensuring USC as long as the conversion efficiency $\mathcal{C}$ is finite.

The GL coefficient $a_y'=\partial_q\alpha_y(q)|_{q\to0}$ giving rise to $\mathcal{C}$ is allowed
when the symmetry is lower than or equal to $C_{2h}$ with an in-plane principal axis.
Some transition-metal dichalcogenides have $D_{3d}$ symmetry for a bilayer~\cite{Liu2015-ia,Manzeli2017-zb},
and superconductivity has been observed in atomically thin MoS$_2$~\cite{Ye2012-cc,Saito2015-aj,Lu2015-uu} and NbSe$_2$~\cite{Frindt1972-il,
Staley2009-ap,
Cao2015-kb,
Xi2015-lw,
Xi2015-lf},
for example.
Bernal-stacked bilayer graphene~\cite{McCann2013-fh,De_la_Barrera2022-ne,Zhou2022-gl,Zhang2023-fi,Dong2023-gy} and Li$_x$ZrNCl~\cite{Kasahara2015-sn,Nakagawa2018-cw,Nakagawa2021-ka}
also have $D_{3d}$-symmetric crystal structures and show superconductivity.
Since the point group $D_{3d}$ recasts into $C_{2h}$ by suppressing threefold-rotation symmetry,
these materials, e.g., under uniaxial strain~\cite{Manzeli2017-zb} or with nematic orders~\cite{Zhou2022-gl,Zhang2023-fi,Dong2023-gy}, are the experimental platform of the 
USC.
To obtain a large conversion efficiency ${\mathcal{C}}$,
strong-coupling superconductors with small $\xi_0$ such as twisted graphene multilayers~\cite{Cao2018-hg,Park2021-kk,Hao2021-ux,Park2022-do} and heavy-fermion superlattices~\cite{Mizukami2011-jx,Naritsuka2021-ym},
as well as systems near a pair-density-wave transition with small $|a_x|$~\cite{Fischer2023-rf,Khim2021-er}, are suitable once the symmetry requirement is satisfied.

The actual current-resistivity relation may differ from our results for various reasons.
While $j_{\rm c-}$ in Fig.~\ref{fig:jq_ss} is increased by the electric field, this effect would be difficult to see when resistive transitions are triggered by the depinning of vortices, phase-slippage events, and so on, depending on sample quality and dimensionality.
This would also reduce the dissipation-induced nonreciprocity from the ideal one [Eq.~\eqref{eq:eta_low_E}].
Even in such cases, the reduction of $j_{\rm c+}$, including USC, is determined by our mechanism under sufficiently large electric fields.
While we have concentrated on the Cooper-pair transport, the normal quasiparticles may also contribute to the in-plane transport since the perpendicular direction is not a principal axis in our system.
This effect may make the onset current of USC finite as in Fig.~\ref{fig:USC}(c), which vanishes in Fig.~\ref{fig:jq_ss}, and will be discussed elsewhere in detail.

\textit{Conclusion.} --- 
In this {paper}, we have proposed a setup to realize USC: applying the out-of-plane electric field and current to a wire or thin film of bilayer superconductors lacking in-plane inversion symmetry.
The inter-layer phase difference $\theta_{\mathcal{E}}$ driven by the out-of-plane electric field is converted into the in-plane Cooper-pair momentum via a GL coefficient $a_y'=\partial_q\alpha_y(q)|_{q\to0}$.
This leads to a one-way current-momentum relation and USC. 
Our result uncovers dissipation-induced SDE free of the external magnetic field with up to $100\%$ diode efficiency.
The precise relation of our results to USC in twisted trilayer graphene~\cite{Lin2022-cz} is left unclear, but
dissipation could be important as illustrated in this {paper}.

\begin{acknowledgments}
We thank Hikaru Watanabe and Kazuaki Takasan for their helpful discussions.
This work was supported by JSPS KAKENHI (Grants No. JP21K13880, No. JP21K18145, No. JP22H01181, No. JP22H04476, No. JP22H04933, No. JP23K17353, and No. JP24H00007).
\end{acknowledgments}

%


\section{Symmetry requirements of $\alpha_y(q)$}
In this section, we discuss the symmetry requirements to obtain a finite $a_y'=\partial_q\alpha_y(q)|_{q\to0}$.
The Pauli matrix $\sigma_y$ can be written as
\begin{align}
\sigma_y&\sim -i\ket{z=d/2}\bra{z=-d/2}\notag\\
&\qquad\qquad+i\ket{z=-d/2}\bra{z=d/2},
\end{align}
which is symmetry-equivalent to the momentum operator in the $z$ direction, as is clear by formally expanding it by $d$.
Thus, $a_y'$ has the same symmetry as 
\begin{align}
a_y'\sim k_xk_z,
\end{align}
which must belong to the identity representation of the point group.
The rotation axis can exist only in the $y$ direction, and the invariance of the bilayer system allows only the twofold rotation.
Thus, the candidate point groups are monoclinic ones and $C_{2v}$, but $k_xk_z$ is prohibited in $C_{2v}$ by the mirror planes.
Thus, the highest-symmetry point group allowing $a_y'$ is $C_{2h}$ with the principal axis along the $y$ direction.
The subgroups $C_1$, $C_i$, $C_s$, $C_2$ also allow $a_y'$.

\section{The bilayer TDGL model}
\subsection{Bilayer TDGL model under electromagnetic fields}
Here we introduce the electric field to the bilayer TDGL equation.
The TDGL equation is given by
\begin{align}
\!\!\!\!\!\!\!\!\frac{\Gamma}{N_0}\partial_t\bm{\psi}(x,t)=-[\halpha(\nabla/i)+\beta_0\bm{\psi}^\dagger(x,t)\bm{\psi}(x,t)]\bm{\psi}(x,t),
\end{align}
in the absence of electromagnetic fields.
By gauge transformation, the order parameter transforms by
\begin{align}
{\psi}'_i(x,t)=e^{-2i\chi_i(x,t)}\psi_i(x,t),
\end{align}
or equivalently,
\begin{align}
\bm{\psi}'(x,t)=e^{-2i\bar{\chi}(x,t)}e^{-i\delta\chi(x,t)\sigma_z}\bm{\psi}(x,t),
\end{align}
with
\begin{subequations}\begin{align}
\bar{\chi}(x,t)&\equiv \frac{\chi_1(x,t)+\chi_2(x,t)}{2},\\
\delta{\chi}(x,t)&\equiv {\chi_1(x,t)-\chi_2(x,t)}.
\end{align}\end{subequations}
The transformed order parameter satisfies
\begin{align}
&\frac{\Gamma}{N_0}[\partial_t+2i\partial_t\bar{\chi}(x,t)+i\partial_t\delta\chi(x,t)\sigma_z]\psi_i'(x,t)\notag\\
&=-e^{-i\delta\chi(x,t)\sigma_z}[\halpha(\nabla/i+2\nabla\bar{\chi})\notag\\
&\qquad+\beta_0\bm{\psi}'^\dagger(x,t)\bm{\psi}'(x,t)]e^{i\delta\chi(x,t)\sigma_z}\bm{\psi}'(x,t).
\end{align}
The electromagnetic field should be incorporated into the equation to cancel the terms dependent on $\chi_i(x,t)$.
Note that vector and scalar potentials $\bm{A}(x,z,t)$ and $\phi(x,z,t)$ satisfy
\begin{subequations}\begin{align}
\delta A(x,t)&\equiv \int_{-d/2}^{d/2} dz\,A_z(x,z,t)\notag\\
\to \delta A'(x,t)&=\int_{-d/2}^{d/2} dz\,A_z(x,z,t)-\partial_z\chi(x,z,t)\notag\\
&=\delta A(x,t)-\delta\chi(x,t),\\
\bar{A}(x,t)&\equiv \frac{A_x(x,d/2,t)+A_x(x,-d/2,t)}{2}\notag\\
\to \bar{A}'(x,t)&=\bar{A}(x,t)-\nabla\bar{\chi}(x,t),\\
\delta \phi(x,t)&\equiv \phi(x,d/2,t)-\phi(x,-d/2,t)\notag\\
\to \delta\phi'(x,t)&=\delta\phi(x,t)+\partial_t\delta\chi(x,t),\\
\bar{\phi}(x,t)&\equiv \frac{\phi(x,d/2,t)+\phi(x,-d/2,t)}{2}\notag\\
\to \bar{\phi}'(x,t)&=\bar{\phi}(x,t)+\partial_t\bar{\chi}(x,t),
\end{align}\end{subequations}
by using $\chi_1(x,t)=\chi(x,d/2,t)$ and $\chi_2(x,t)=\chi(x,-d/2,t)$.
The minimal coupling of the electromagnetic field to our model is thus obtained by formally seting $\bm{A}=0$ and $\phi=0$ and replacing $\bar{\chi}$ and $\delta\chi$ by $\bm{A}'$ and $\phi'$,
\begin{align}
&\frac{\Gamma}{N_0}\left[\partial_t+2i\bar{\phi}(x,t)+i\delta\phi(x,t)\sigma_z\right]\bm{\psi}(x,t)\notag\\
&=-e^{i\delta A(x,t)\sigma_z}[\halpha(\nabla/i-2\bar{A}(x,t))\notag\\
&\qquad+\beta_0\bm{\psi}^\dagger(x,t)\bm{\psi}(x,t)]e^{-i\delta A(x,t)\sigma_z}\bm{\psi}(x,t),
\end{align}
where we removed the prime of the quantities.

We adopt the gauge where the vector potential vanishes and describe the perpendicular electric field by the scalar potential $\phi(z)=-E_\perp z$.
In this case, we obtain $\delta\phi=-E_\perp d$ and 
the TDGL equation recasts into
\begin{align}
&\frac{\Gamma}{N_0}\left[\partial_t-iE_\perp d\sigma_z\right]\bm{\psi}(x,t)\notag\\
&=-[\halpha(\nabla/i)+\beta_0\bm{\psi}^\dagger(x,t)\bm{\psi}(x,t)]\bm{\psi}(x,t),\label{eq:TDGL_SM}
\end{align}
which gives
\begin{align}
&\left[\partial_t+i\Phi\right]\bm{\psi}(q,t)=-[\halpha(q)+\beta_0|\bm{\psi}(q,t)|^2]\bm{\psi}(q,t),
\end{align}
with $t\to \Gamma t/N_0$ and $\Phi=-\Gamma E_\perp d\sigma_z/N_0$, for a plain-wave order parameter $\bm{\psi}(x,t)=e^{iqx}\bm{\psi}(q,t)$.

When the system is in equilibrium, we can obtain the coupling of electromagnetic fields to the GL free energy in the same way:
\begin{align}
F[\bm{\psi}]&=N_0\int dx\,\bm{\psi}^\dagger(x)e^{i\delta A(x)\sigma_z}\halpha(\nabla/i-2\bar{A}(x))\notag\\
&\qquad\cdot e^{-i\delta A(x)\sigma_z}\bm{\psi}(x)+\frac{\beta_0}{2}[\bm{\psi}^\dagger(x)\bm{\psi}(x)]^2.
\end{align}
The current in the $x$ direction at the position $x$ is obtained by
\begin{align}
j_\parallel(x)&=-\int_{-d/2}^{d/2} dz\,\frac{\delta F}{\delta A_x(x,z)}
=-\frac{\delta F}{\delta \bar{A}(x)},\label{eq_SM:supercurrent}
\end{align}
while the current density in the $z$ direction is obtained by
\begin{align}
j_\perp(x)&=-\frac{1}{d}\int_{-d/2}^{d/2}dz\,\frac{\delta F}{\delta A_z(x,z)}=-\frac{\delta F}{\delta [\delta A(x)]}.
\end{align}
We assume that these formulas give electric current also in nonequilibrium by substituting for $\bm{\psi}(x)$ the solution of the TDGL equation.
This is a standard prescription at least for the well-studied case, i.e., $j_\parallel(x)$ in the absence of the layer degree of freedom~\cite{Rubinstein2007-ck, Chtchelkatchev2012-jd,Kopnin2001-ie}.
For the plain-wave order parameter $\bm{\psi}(x,t)=e^{iqx}\bm{\psi}(q,t)$, the electric currents are evaluated by
\begin{subequations}\begin{align}
j_\parallel(x,t)&=j_\parallel(q)=2N_0\bm{\psi}^\dagger(q,t)\partial_q\halpha(q)\bm{\psi}(q,t),\\
j_\perp(x,t)&=j_\perp(q)=-iN_0\bm{\psi}^\dagger(q,t)[\sigma_z,\halpha(q)]\bm{\psi}(q,t),
\end{align}\label{eq:currents_SM}\end{subequations}
for our gauge $\bar{A}(x)=\delta A(x)=0$.
In this gauge, the time dependence of the electric current, if any, comes only from $\bm{\psi}(x,t)$.

\subsection{The steady states}

The TDGL equation~\eqref{eq:TDGL_SM} is rewritten as the coupled equations as follows.
By writing $\bm{\psi}(q,t)=e^{i\chi(q,t)}R(q,t)\hat{\psi}(q,t)$ with $\chi(q,t)\in\mathbb{R}$, $R(q,t)\ge0$ and $|\hat{\psi}(q,t)|=1$,
we obtain
\begin{subequations}\begin{align}
\partial_t\chi(q,t)&=-\braket{{\Phi}}_{\hat{\psi}(q,t)}+i\hat{\psi}(q,t)^\dagger\partial_t\hat{\psi}(q,t),\\
\partial_tR(q,t)&=-(\braket{{\hat{\alpha}(q)}}_{\hat{\psi}(q,t)}+\beta_0 R^2(q,t))R(q,t),\\
(1-\hat{\psi}(q,t)&\hat{\psi}(q,t)^\dagger)\partial_t\hat{\psi}(q,t)\notag\\
&=-\left[\hat{\alpha}_{\rm eff}(q)-\braket{{\hat{\alpha}_{\rm eff}(q)}}_{\hat{\psi}(q,t)}\right]\hat{\psi}(q,t),
\end{align}\end{subequations}
with $\braket{O}_{\hat{\psi}(q,t)}\equiv\hat{\psi}^\dagger(q,t) O\hat{\psi}(q,t)$, $\hat{\alpha}_{\rm eff}(q)\equiv\hat{\alpha}(q)+i\Phi$, and $\Phi=-\mE\sigma_z$.
The first and second lines are obtained by acting $\hat{\psi}^\dagger(q,t)$ from the left and taking the real and imaginary parts of the equation.
The third line is obtained by plugging the first and second lines into the original equation.
The first equation is the Josephson equation, relating the evolution of the overall phase of the superconductor with the averaged scalar potential.
The second and third equations describe the dynamics of the amplitude and internal degrees of freedom of the order parameter, respectively.

The steady-state solution of the TDGL equation is given by setting $\partial_t\hat{\psi}(q,t)=0$ and $\partial_tR(q,t)=0$.
We obtain
\begin{align}
\alphaeff(q)\hat{\psi}_\sst(q)&=\braket{\alphaeff(q)}_{\hat{\psi}_\sst(q)}\hat{\psi}_\sst(q),
\end{align}
which means that $\hat{\psi}_\sst(q)$ is the normalized eigenstate of $\alphaeff(q)$.
We also obtain
\begin{align}
R^2_\sst(q)=-\frac{\braket{\alpha(q)}_{\hat{\psi}_\sst(q)}}{\beta_0}.
\end{align}
Thus, we need to diagonalize $\alphaeff(q)$ and obtain its eigenstates to proceed further, which is discussed in the next section along with the derivation of the current-momentum relation.

\subsection{Diagonalization of $\alphaeff(q)$}
In this section, we derive the normalized eigenstates $\ket{\pm(q)}$ satisfying $\alphaeff(q)\ket{\pm(q)}=\lambda_\pm(q)\ket{\pm(q)}$ for
\begin{align}
    \alphaeff(q)&=\alpha_0(q)+\alpha_x(q)\sigma_x+\alpha_y(q)\sigma_y-i\mE\sigma_z\\
    &=e^{-\frac{i}{2}\alpha_\theta(q)\sigma_z}[\alpha_0(q)+\alpha_\rho(q)\sigma_x-i\mE\sigma_z]e^{\frac{i}{2}\alpha_\theta(q)\sigma_z}.\notag
\end{align}
To obtain the second line, we used
\begin{align}
\alpha_x(q)&=\alpha_\rho(q)\cos\alpha_\theta(q),\  \alpha_y(q)=\alpha_\rho(q)\sin\alpha_\theta(q).
\end{align}

To diagonalize $\alphaeff(q)$, we first consider the diagonalization of the matrix of the form $\halpha'(q)=\sigma_x-i\sin\theta_\mE(q)\sigma_z$, with the angle $\theta_\mE(q)$ introduced by
\begin{align}
    \sin\theta_\mE(q)\equiv\frac{\mE}{\alpha_\rho(q)},\quad\cos\theta_\mE(q)>0.
\end{align}
The matrix $\halpha'(q)$ satisfies $\halpha'(q)^2=1-\sin^2\theta_\mE(q)=\cos^2\theta_\mE(q)$, and thus has eigenvalues $\pm\cos\theta_\mE(q)$.
Since $(\halpha'(q)+\cos\theta_\mE(q))(\halpha'(q)-\cos\theta_\mE(q))=0$, the eigenstate of $\halpha'(q)$ with the eigenvalue $-\cos\theta_\mE(q)<0$ is obtained by a column vector of
\begin{align}
\halpha'(q)-\cos\theta_\mE(q)
    &=\begin{pmatrix}
    -e^{i\theta_\mE(q)}&1\\1&-e^{-i\theta_\mE(q)}
    \end{pmatrix},
\end{align}
while that for $\cos\theta_\mE(q)$ is given by a column vector of
\begin{align}
    \halpha'(q)+\cos\theta_\mE(q)&=\begin{pmatrix}
        e^{-i\theta_\mE(q)}&1\\1&e^{i\theta_\mE(q)}
    \end{pmatrix}.
\end{align}
These eigenstates of $\halpha'(q)$ give those of $\alphaeff(q)$ after the unitary transformation $e^{\frac{i}{2}\alpha_\theta(q)\sigma_z}$, and thus we obtain 
the normalized eigenstates of $\alphaeff(q)$, 
\begin{align}
    \hat{\psi}_{\rm ss}(q)&=\ket{-(q)}\equiv\frac{1}{\sqrt{2}}e^{-\frac{i}{2}[\alpha_\theta(q)-\theta_\mE(q)]\sigma_z}\begin{pmatrix}
        1\\-1
    \end{pmatrix},\label{eq:ss_SM}
\end{align}
and 
\begin{align}
    \ket{+(q)}&=\frac{i}{\sqrt{2}}e^{-\frac{i}{2}[\alpha_\theta(q)+\theta_\mE(q)]\sigma_z/2}\begin{pmatrix}
        1\\1
    \end{pmatrix},
\end{align}
corresponding to the eigenvalues of $\alphaeff(q)$,
\begin{align}
    \lambda_\pm(q)&=\alpha_0(q)\pm\alpha_\rho(q)\cos\theta_\mE(q)\notag\\
    &=\alpha_0(q)\pm\sqrt{\alpha_\rho(q)^2-\mE^2},
\end{align}
respectively.
The overall phase $i$ of $\ket{+(q)}$ is chosen to make $\braket{-(q)|+(q)}$ to be real for convenience:
\begin{align}
\braket{-(q)|+(q)}&=\frac{i}{2}(1,-1)e^{-i\theta_\mE(q)\sigma_z}\begin{pmatrix}
    1\\1
\end{pmatrix}\notag\\
&=\sin\theta_\mE(q).
\end{align}
This is finite in the presence of the perpendicular electric field as a result of the non-Hermitian property $\alphaeff(q)\neq \alphaeff^\dagger(q)$.

According to the TDGL equation, the amplitude $R_\sst(q)$ of the steady state satisfies
\begin{align}
R^2_\sst(q)&=-\frac{\hat{\psi}^\dagger_{\rm ss}(q)\halpha(q)\hat{\psi}_{\rm ss}(q)}{\beta_0}\notag\\
&=-\frac{1}{\beta_0}\Re[\braket{-(q)|\alphaeff(q)|-(q)}]\notag\\
&=-\frac{1}{\beta_0}\Re[\lambda_-(q)].
\end{align}
Here, the eigenvalue is expanded as
\begin{align}
\lambda_-(q)&=\alpha_0(q)-\sqrt{\alpha_\rho(q)^2-\mE^2}\notag\\
&\simeq \alpha_0(q)-\alpha_\rho(q)+\frac{\mE^2}{2\alpha_\rho(q)}\notag\\
&\simeq\epsilon+\xi_0^2q^2+\frac{\mE^2}{2|a_x|}.\label{eq:lambdaminus_SM}
\end{align}
Here, we approximated the expression by assuming $\mE$ and $q$ are small quantities.
This assumption is consistent since both $q$ and $\mE$ must be at most $O(\sqrt{|\epsilon|})$ for $R_\sst^2(q)$ to be positive, as is clear from the last line of Eq.~\eqref{eq:lambdaminus_SM}, where only $\epsilon=-|\epsilon|$ is a negative quantity.
Note that $|a_x|=|\alpha_x(0)|$ is assumed not to be small, which is always satisfied for the temperature $T$ sufficiently near $T_{\rm c}$ unless $|a_x|$ accidentally vanishes (corresponding to the pair-density-wave transition point~\cite{Fischer2023-rf}).
At the same level of the approximation, we obtain
\begin{align}
\theta_\mE(q)&\simeq\frac{\mE}{|a_x|},\quad 
\alpha_\theta(q)\simeq\alpha_\theta(0)+\frac{a_y'q}{a_x},
\end{align}
which gives the expression of $\hat{\psi}_\sst(q)$ in the main text.
The equilibrium properties of the system are reproduced by setting $\mE=0$.

\subsection{Evaluation of the electric current}
The electric currents can be evaluated by 
plugging the expression of the steady state Eq.~\eqref{eq:ss_SM} into Eqs.~\eqref{eq:currents_SM}.
The electric current through the system is calculated by
\begin{align}
j_\parallel(q)&=2N_0R_\sst^2(q)\braket{-(q)|\partial_q\halpha(q)|-(q)}\notag\\
&=2N_0R_\sst^2(q)[2a_0''q+a_y'\braket{-(q)|\sigma_y|-(q)}]\notag\\
&=2N_0R_\sst^2(q)[2a_0''q-a_y'\sin(\alpha_\theta(q)-\theta_\mE(q))].
\end{align}
This is approximated by
\begin{align}
j_\parallel(q)&\simeq 2N_0R_\sst^2(q)\left[2a_0''-a_y'\left(\frac{a_yq}{|a_x|}-\frac{a_x}{|a_x|}\theta_\mE\right)\right]\notag\\
&=2N_0R_\sst^2(q)\left[2\xi_0^2q+\frac{a_y'}{a_x}\mE\right]\notag\\
&=\frac{4N_0\xi_0^2}{\beta_0}[-\bar{\epsilon}(\mE)-\xi_0^2q^2](q-q_\mE).
\end{align}

The expression of $j_\parallel(q)$ can be reproduced by focusing on the momentum shift in the steady-state solution, i.e., $\hat{\psi}_{\rm ss}(q)\simeq\hat{\psi}_{\rm eq}(q-\tilde{q}_\mE)$ with $\tilde{q}_\mE\equiv a_x\theta_\mE/a_y'$.
By using $\partial_q\hat{\alpha}(q)=2a_0''q+a_y'\sigma_y$, we obtain
\begin{align}
    &\hat{\psi}_{\rm ss}^\dagger(q)\partial_q\hat{\alpha}(q)\hat{\psi}_{\rm ss}(q)\notag\\
    &=\hat{\psi}_{\rm eq}^\dagger(q-\tilde{q}_\mE)[\partial_q\hat{\alpha}(q)]_{q\to q-\tilde{q}_\mE}\hat{\psi}_{\rm eq}(q-\tilde{q}_\mE)\notag\\
    &\quad+\hat{\psi}_{\rm ss}^\dagger(q)[\partial_q\hat{\alpha}(q)-[\partial_q\hat{\alpha}(q)]_{q\to q-\tilde{q}_\mE}]\hat{\psi}_{\rm ss}(q)\notag\\
    &=2\xi_0^2(q-\tilde{q}_\mE)+2a_0''\tilde{q}_\mE\notag\\
    &=2\xi_0^2\left(q-\left[1-\frac{a_0''}{\xi_0^2}\right]\tilde{q}_\mE\right).
\end{align}
This reproduces the previous result by noting that
\begin{align}
    \left[1-\frac{a_0''}{\xi_0^2}\right]\tilde{q}_\mE&=\frac{(a_0''-a_y'^2/2|a_x|)-a_0''}{\xi_0^2}\left(\frac{a_x}{a_y'}\theta_\mE\right)\notag\\
    &=-\frac{a_y'a_x}{2\xi_0^2|a_x|}\theta_\mE.
\end{align}

Another way to understand $j_\parallel(q)$ is to use the left eigenstates $\rket{\pm(q)}$ of $\alphaeff(q)$, which is defined by $\rbra{\pm(q)}\alphaeff(q)=\rbra{\pm(q)}\lambda_\pm(q)$ and is introduced later in detail.
By using $\partial_q\halpha=\partial_q\alphaeff$, we can also write
\begin{align}
    \hat{\psi}_{\rm ss}^\dagger\partial_q\halpha\hat{\psi}_{\rm ss}&=\braket{-|\partial_q\halpha|-}\notag\\
    &=\Re[\braket{-|\partial_q\halpha|-}]\notag\\
    &=\Re\Bigl[\rbraket{-|\partial_q\alphaeff|-}+\braket{-|+}\rbraket{+|\partial_q\halpha|-}\Bigr]\notag\\
    &=\partial_q\Re[\lambda_-]+\Re\left[\braket{-|+}\rbraket{+|\partial_q\halpha|-}\right],
\end{align}
where the argument $q$ is abbreviated.
The first term corresponds to $2\xi_0^2q$ and thus the deviation from it results from the nonorthogonality of the (right) eigenstates $\braket{-(q)|+(q)}\neq0$, i.e., the non-Hermitian nature of the matrix $\alphaeff(q)$.

The perpendicular electric current is evaluated by 
\begin{align}
j_\perp(q)&=-iN_0R^2_\sst(q)\braket{-(q)|[\sigma_z,\halpha(q)]|-(q)}\notag\\
&=2N_0R^2_\sst(q)\braket{-(q)|\alpha_x(q)\sigma_y-\alpha_y(q)\sigma_x|-(q)}\notag\\
&=2N_0R_\sst^2(q)\Bigl[-\alpha_x(q)\sin(\alpha_\theta(q)-\theta_\mE(q))\notag\\&\qquad\qquad+\alpha_y(q)\cos(\alpha_\theta(q)-\theta_\mE(q))\Bigr]\notag\\
&=2N_0R_\sst^2(q)\alpha_\rho(q)\sin\theta_\mE(q).
\label{eq:ACJJ_like}
\end{align}
It should be noted that a finite voltage is required for a finite Cooper-pair-hopping current because a strongly-coupled bilayer is considered rather than weak links.
Note also that $0=d\theta_\mE/dt\neq E_\perp d$ and Eq.~\eqref{eq:ACJJ_like} is different from the a.c. Josephson relation.
Indeed, the Josephson relation holds only for superconductors in equilibrium, i.e., in the absence of the gauge-invariant scalar potential~\cite{Kopnin2001-ie}, which corresponds to $\Phi$ for the steady state of our model.
A finite gauge-invariant scalar potential implies that the conversion of the Cooper-pair current to the normal current takes place~\cite{Kopnin2001-ie}, and thus 
the net perpendicular current in (and around) the system would generally be contributed by the normal current in addition to the perpendicular flow of Cooper pairs evaluated by Eq.~\eqref{eq:ACJJ_like}.

\section{Phase diagram for a given current-momentum relation}
In this section, we discuss the phase diagram of the unidirectional superconductivity for a given current-momentum relation.
After a scale transformation of $j_\sst(q)$ in the main text, we obtain
\begin{align}
j_{\sst}(Q)=\frac{3\sqrt{3}}{2}j_{\rm c0}(Q-Q_\mE)(1-Q^2),\quad |Q|\le1
\label{eq:current-momentum-scale}
\end{align}
with
\begin{align}
Q=\xi q,\quad \xi=\frac{\xi_0}{\sqrt{-\bar{\epsilon}(\mE)}},\quad \bar{\epsilon}(\mE)=\epsilon\left(1-\frac{\mE^2}{\mE_{\rm c}(\epsilon)^2}\right),
\end{align}
and
\begin{align}
Q_\mE=\xi q_\mE=\frac{\xi_0}{\sqrt{-\bar{\epsilon}(\mE)}}\frac{dq_\mE}{d\mE}\mE\equiv
{\mC}\frac{{\mE}/{\mE_{\rm c}(\epsilon)}}{\sqrt{1-[{\mE}/{\mE_{\rm c}(\epsilon)]^2}}}.
\end{align}
Here $\mE_{\rm c}(\epsilon)=\mE_{\rm c}(-1)\sqrt{-\epsilon}$ and we defined
\begin{align}
\mC\equiv \xi_0\frac{dq_\mE}{d\mE}\mE_{\rm c}(-1).
\end{align}
The constant $\mC$ for our model is given by
\begin{align}
    \mC&=\xi_0\frac{-a_y'}{2\xi_0^2a_x}\sqrt{2|a_x|}\notag\\
    &=\sgn[-a_y'a_x]\frac{|\xi_0^{-1}a_y'|}{\sqrt{2|a_x|}}.
\end{align}

We assume that $\mE>0$ and $\mC>0$ without loss of generality, which corresponds to the case of Fig.~3 in the main text.
Then, the onset of the unidirectional superconductivity is given by the condition $\frac{dj_{\rm ss}}{dQ}(1)=0$.
This is equivalent to $Q_\mE=1$, namely $\mE=\mE_{\rm c2}(\epsilon)$ given by the ratio
\begin{align}
\frac{\mE_{\rm c2}(\epsilon)}{\mE_{\rm c}(\epsilon)}&=\frac{1}{\sqrt{1+\mC^2}},
\end{align}
which is independent of $\epsilon$.
The depairing critical current for the positive direction is given by
\begin{align}
\jcp
&=\max_{|Q|\le 1} j_{\rm c0}(Q),
\end{align}
which vanishes for $\mE>\mE_{\rm c2}(\epsilon)$.
The depairing critical current for the negative direction is given by
\begin{align}
\jcm&=\min_{|Q|\le 1} j_{\rm c0}(Q).
\end{align}
The extrema of $j_\sst(Q)$ are obtained at $Q=Q_{\rm c\pm}$ such that
\begin{align}
\left.\frac{dj_\sst(Q)}{dQ}\right|_{Q=Q_{\rm c\pm}}=0,\quad Q_{\rm c+}>Q_{\rm c-}.
\end{align}
The solutions are 
\begin{align}
Q_{\rm c\pm}&=\frac{Q_\mE\pm\sqrt{Q_\mE^2+3}}{3},
\end{align}
and thus we obtain with the Heaviside step function $\theta(x)$,
\begin{align}
\jcp=j_\sst(\min[1,Q_{\rm c+}]),\quad
\jcm=|j_\sst(Q_{\rm c-})|.
\end{align}
The diode quality factor is then given by
\begin{align}
\eta\equiv\frac{\jcp-\jcm}{\jcp+\jcm}.
\end{align}
Figures~4(a) and~(b) in the main text are obtained by plotting the above results in terms of $\epsilon$ and/or $\mE$.
The quality factor is linear when $|\mE|$ is sufficiently small and then is explicitly given by
\begin{align}
\eta=-\sqrt{3}\mC\frac{\mE}{\mE_{\rm c}(\epsilon)}+O(\mE^2).
\end{align}

\section{Local stability of the steady state without the current bias}
In this section, we discuss the local stability of the steady state following Refs.~\cite{Langer1967-rh,Samokhin2017-xf},
for the dynamics determined by
the TDGL equation alone.
Here we do not require that the electric current is constant in time and space, which means the absence of the external current basis.
Therefore, the situation under consideration would physically correspond to persistent-current geometry such as the ring-shaped superconductors.
Local stability condition with the current bias is studied at the end of the Supplemental Material for the general bilayer TDGL model,
where TDGL is solved simultaneously with the fixed-current condition.
It turns out that these two local stability conditions with and without the current bias do not coincide, in stark contrast to the equilibrium case.

The steady state is given by
\begin{align}
    \bm{\psi}_\sst(x,t)=e^{iqx}e^{{i\chi_\sst(q,t)}}\bm{\psi}_\sst(q),
\end{align}
with $\bm{\psi}_\sst(q)=R_\sst(q)\hat{\psi}_\sst(q)$ and
\begin{subequations}\begin{align}
    {\partial_t\chi_\sst(q,t)}&=-\Im[\lambda_-(q)],\\
    \beta_0R^2_\sst(q)&=-\Re[\lambda_-(q)],\\
    \alphaeff(q)\hat{\psi}_\sst(q)&=\lambda_-(q)\hat{\psi}_\sst(q).
\end{align}\end{subequations}
Here, $\lambda_-(q)$ is the eigenvallue of $\alphaeff(q)$ with the smallest real part and is assumed to satisfy $\Re[\lambda_-(q)]< 0$.
The first line follows from
\begin{align}
\partial_t\chi_\sst(q,t)&=-\hat{\psi}_\sst^\dagger(q)\Phi\hat{\psi}_\sst(q)\notag\\
&=-\Im[\hat{\psi}_\sst^\dagger(q)\alphaeff(q)\hat{\psi}_\sst(q)]\notag\\
&=-\Im[\lambda_-(q)].
\end{align}

We are interested in whether an infinitesimal deviation $\delta\bm{\psi}(x,t)$ from $\bm{\psi}_\sst(x,t)$ as given by 
\begin{align}
    \bm{\psi}'(x,t)=e^{iqx}e^{i\chi_\sst(t)}[\bm{\psi}_\sst(q)+\delta\bm{\psi}(x,t)],
\end{align}
decays in time.
To see this, we linearize the TDGL equation
\begin{align}
    \partial_t\bm{\psi}'(x,t)&=-\Bigl[\halpha(\nabla/i)+i\Phi+\beta_0|\bm{\psi}'(x,t)|^2\Bigr]\bm{\psi}'(x,t),
\end{align}
in terms of $\delta\bm{\psi}(x,t)$ and obtain
\begin{align}
    \partial_t\delta\bm{\psi}(x,t)&=-\Bigl[\halpha(q+\nabla/i)+i\Phi+\beta_0|\bm{\psi}_\sst(q)|^2\notag\\
    &\qquad+i\partial_t\chi_\sst(q,t)+\beta_0\bm{\psi}_\sst(q)\bm{\psi}_\sst(q)^\dagger\Bigr]\delta\bm{\psi}(x,t)\notag\\
    &\qquad-\Bigl[\beta_0\bm{\psi}_\sst(q)\bm{\psi}_\sst^T(q)\Bigr]\delta\bm{\psi}^*(x,t).
\end{align}
By switching to the momentum representation
\begin{align}
    \delta\bm{\psi}(x,t)&=\int\frac{dk}{\sqrt{2\pi}}\delta\bm{\psi}(k,t)e^{ikx},
\end{align}
we obtain
\begin{align}
    \partial_t\begin{pmatrix}
        \delta\bm{\psi}(k,t)\\
\delta\bm{\psi}^*(-k,t)
    \end{pmatrix}&=-M_k(q)\begin{pmatrix}
        \delta\bm{\psi}(k,t)\\
\delta\bm{\psi}^*(-k,t)
    \end{pmatrix},
\end{align}
with the matrix
\begin{widetext}
\begin{align}
    M_k(q)\equiv\begin{pmatrix}
        \halpha(q+k)+i\Phi-\lambda_-(q)+\beta_0\bm{\psi}_\sst(q)\bm{\psi}_\sst(q)^\dagger&\beta_0\bm{\psi}_\sst(q)\bm{\psi}_\sst(q)^T\\
        \beta_0\bm{\psi}_\sst(q)^*\bm{\psi}_\sst^\dagger(q)&\halpha(q-k)^*-i\Phi^*-\lambda_-(q)^*+\beta_0\bm{\psi}_\sst(q)^*\bm{\psi}_\sst(q)^T
    \end{pmatrix}.
\end{align}
\end{widetext}

The steady state is stable when all the eigenvalues of $M_k(q)$ have a positive real part for arbitrary $k$, with an exception at $k=0$:
It should be noted that $M_{k=0}(q)$ satisfies
\begin{align}
    M_0(q)\begin{pmatrix}
        \bm{\psi}_\sst(q)\\
        -\bm{\psi}_\sst^*(q)
    \end{pmatrix}=0,
\end{align}
and thus has a zero mode.
However, this mode corresponds to the uniform phase rotation $\delta\bm{\psi}(x,t)=i\delta\phi\,\bm{\psi}_\sst(q)$ with a small real parameter $\delta\phi$, and the steady state is physically unchanged.
Thus, we focus on the other eigenvalues of $M_k(q)$.

When the wave number $k$ is large, $M_k(q)$ is dominated by $\alpha_0(q\pm k)\sim a_0''k^2$, and thus all the eigenvalues have a large positive real part.
The stability would thus be determined by the behavior around $k=0$, as actually is the case for equilibrium situations~\cite{Langer1967-rh,Samokhin2017-xf}.
For this purpose, we are interested in the long-wave-length behavior of the branch $m_0(k,q)$ connecting to the zero mode $m_0(0,q)=0$, since the other eigenvalues of $M_{0}(q)$ are positive when $\Re[\lambda_+(q)-\lambda_-(q)]>0$ and $-\Re[\lambda_-(q)]>0$ as we see later.

Note that the matrix $M_k(q)$ satisfies the particle-hole-like symmetry
\begin{align}
\tau_xM_{-k}^*(q)\tau_x=M_k(q).
\end{align}
This ensures $m_0({k},q)^*=m_0({-k},q)$ for small $k$ since $m_0(0,q)$ is not degenerate, and therefore we obtain
\begin{align}
    \left.\frac{d^{2n+1}m_0(k,q)}{dk^{2n+1}}\right|_{k=0}\in i\mathbb{R},\quad \left.\frac{d^{2n}m_0(k,q)}{dk^{2n}}\right|_{k=0}\in \mathbb{R},
\end{align}
as well as the Taylor expansion of the form
\begin{subequations}\begin{gather}
    m_0(k,q)=im_0'(q)k+\frac{k^2}{2}m_0''(q)+O(k^3),\\
m_0'(q),\,m_0''(q)\in\mathbb{R}.
\end{gather}\end{subequations}
The first term (and all the odd-order terms) does not contribute to $\Re[m_0(k,q)]$, and thus the stability problem recasts into the sign problem of
\begin{align}
    m_0''(q)=\left.\frac{\partial^2m_0(k,q)}{\partial k^2}\right|_{k=0},
\end{align}
which is discussed in the subsequent sections.

\subsection{Eigenproblem of $M_0$}
\label{sec:eigenproblemofM0}

In the following sections, we discuss the behavior of the eigenvalues of $M_k$ near $k=0$. 
For this purpose, we solve the eigenproblem of $M_0$ in this section.
We start from the right eigenstates of the non-Hermitian matrix $\alphaeff$,
\begin{align}
\alphaeff\ket{\pm}&=\lambda_\pm\ket{\pm}.
\end{align}
Here and hereafter, the argument $q$ of quantities is abbreviated when it is clear.
We normalize the right eigenstates, $\braket{i|i}=1$ and also choose the phase of $\ket{+}$ to satisfy 
\begin{align}
c\equiv\braket{-|+}\in\mathbb{R},
\label{eq:def_c}
\end{align}
for convenience.

We assume that $\mE$ in $\alphaeff$ is sufficiently small to allow the diagonalization of $\alphaeff$, which is always satisfied for the strength of the electric field we are interested in.
Then, we obtain the diagonalization
\begin{align}
    A^{-1}\alphaeff A=\diag(\lambda_+,\lambda_-),
\end{align}
by using the regular matrix
\begin{align}
    A\equiv(\ket{+}\,\ket{-}).
\end{align}
We define the states $\rbra{\pm}$ by
\begin{align}
A^{-1}\equiv\begin{pmatrix}
        \rbra{+}\\\rbra{-}
    \end{pmatrix}.
\end{align}
The states $\rbra{\pm}$ are the left eigenstates of $\alphaeff$,
\begin{align}
\rbra{\pm}\alphaeff=\rbra{\pm}\lambda_\pm,
\end{align}
since $A^{-1}\alphaeff=\diag(\lambda_+,\lambda_-)A^{-1}$
and satisfy the biorthogonality $\rbraket{i|j}=\delta_{ij}$ since $A^{-1}A=1$.
By defining the projection operators $p_i\equiv\ket{i}\rbra{i}$, we obtain the spectral decomposition of $\alphaeff$
\begin{align}
    \alphaeff=A\diag(\lambda_+,\lambda_-)A^{-1}=\lambda_+p_++\lambda_-p_-,
\end{align}
where $p_ip_j=\delta_{ij}p_i$ and $p_++p_-=1$ but $p_i^\dagger\neq p_i$ in general.
For more mathematical details of non-Hermitian matrices, see Ref.~\cite{Ashida2020-yn}.

With the above notations, we can write
\begin{align}
    \hat{\psi}_\sst=\ket{-}.
\end{align}
Thus, the matrix $M_k$ is given by
\begin{widetext}
\begin{align}
    M_k&=\begin{pmatrix}
        \alphaeff(q+k)-\lambda_--\Re[\lambda_-]\ket{-}\bra{-}&-\Re[\lambda_-]\ket{-}\bra{-^*}\\
        -\Re[\lambda_-]\ket{-^*}\bra{-}&\alphaeff(q-k)^*-\lambda_-^*-\Re[\lambda_-]\ket{-^*}\bra{-^*}
    \end{pmatrix}.
\end{align}    
By using $\Delta\lambda=\lambda_+-\lambda_-$ and $\tilde{A}=\diag(A,A^*)$, we obtain the transformation
\begin{align}
    \tilde{M}_0&\equiv\tilde{A}^{-1}M_0\tilde{A}\notag\\
    &=\begin{pmatrix}\Delta\lambda \frac{1+\sigma_z}{2}-\Re[\lambda_-](\frac{1-\sigma_z}{2}+c\sigma_-)&-\Re[\lambda_-](\frac{1-\sigma_z}{2}+c\sigma_-)\notag\\
    -\Re[\lambda_-](\frac{1-\sigma_z}{2}+c\sigma_-)&\Delta\lambda^*\frac{1+\sigma_z}{2}-\Re[\lambda_-](\frac{1-\sigma_z}{2}+c\sigma_-)
    \end{pmatrix}\\
    &=\left(\Re[\Delta\lambda]+i\tau_z\Im[\Delta\lambda]\right)\frac{1+\sigma_z}{2}-\Re[\lambda_-](1+\tau_x)\left(\frac{1-\sigma_z}{2}+c\sigma_-\right).
\end{align}
\end{widetext}
Here, $\tau_i$ indicates the Pauli matrices in the doubling space and $\sigma_-=(\sigma_x-i\sigma_y)/2$.
By using \textit{Mathematica}, the spectral decomposition of $\tilde{M}_0$ is given by
\begin{align}
    \tilde{M}_0=0\cdot \tilde{P_0}-2\Re[\lambda_-]\tilde{P_1}+\Delta\lambda \tilde{P_2}+\Delta\lambda^*\tilde{P_3},
\end{align}
with projection operators
\begin{subequations}
    \begin{align}
    \tilde{P_0}&=\frac{1-\tau_x}{2}\frac{1-\sigma_z}{2},\\ \tilde{P_1}&=\frac{1+\tau_x}{2}\frac{1-\sigma_z}{2}-\sigma_-\begin{pmatrix}
    C&C^*\\
    C&C^*
    \end{pmatrix}_\tau,\\
    \tilde{P_2}&=\frac{1+\tau_z}{2}\frac{1+\sigma_z}{2}+\sigma_-\begin{pmatrix}
        C&0\\
        C&0
    \end{pmatrix}_\tau,\\
    \tilde{P_3}&=\frac{1-\tau_z}{2}\frac{1+\sigma_z}{2}+\sigma_-\begin{pmatrix}
        0&C^*\\
        0&C^*
    \end{pmatrix}_\tau,
\end{align}
\end{subequations}
for eigenvalues
\begin{align}
    m_0=0,\  m_1=-2\Re[\lambda_-], m_2=\Delta\lambda,\ m_3=\Delta\lambda^*,
\end{align}
respectively, which satisfy $\tilde{P}_i\tilde{P}_j=\delta_{ij}\tilde{P}_i$ as well as $\sum_i\tilde{P}_i=1$.
The subscript $\tau$ indicates matrices in the doubling space.
Here, we defined a constant
\begin{align}
    C&=-\frac{c\,\Re[\lambda_-]}{2\Re[\lambda_-]+\Delta\lambda},
\end{align}
with $c$ defined in Eq.~\eqref{eq:def_c}.

The matrix $\tilde{M}_0$, and thus $M_0$, have two real eigenvalues $0$ and $-2\Re[\lambda_-]>0$ corresponding to the Nambu-Goldstone mode and Higgs mode as well as a pair of eigenvalues $\Delta\lambda$ and $\Delta\lambda^*$.
Since $\Re[\Delta\lambda]>0$ is satisfied by assumption, the steady state is stable against spatially-uniform perturbation.

\subsection{Evaluation of $m_0''$}

The quantity $m_0''$ can be evaluated with the formula
\begin{align}
    m_0''&=[\partial_k^2m_0(k)]_{k=0}\notag\\
    &=\Tr[\partial_k^2M_k{P_0}]+2\sum_{i\neq 0}\frac{\Tr[\partial_kM_k P_i\partial_kM_k{P_0}]}{m_0-m_i},\label{eq:formula_Hel}
\end{align}
whose derivation is given in the next subsection.
Here, $P_i=\tilde{A}\tilde{P}_i\tilde{A}^{-1}$ represents the projection operators of $M_0$.
$\partial_kM_k$ and $\partial_k^2M_k$ indicate those at $k=0$, and are given by
\begin{align}
    \!\!\!\!\partial_kM_k&=\begin{pmatrix}
        \partial_q\halpha&0\\0&-\partial_q\halpha^*
    \end{pmatrix},\ \partial_k^2M_k&=\begin{pmatrix}
        \partial_q^2\halpha&0\\0&\partial_q^2\halpha^*
    \end{pmatrix}.
\end{align}
Let us define
\begin{subequations}\begin{align}
\tilde{V}&\equiv \tilde{A}^{-1}\partial_kM_k\tilde{A}=\diag(\tilde{v},-\tilde{v}^*),\\
\tilde{v}&\equiv A^{-1}\partial_q\halpha A=\begin{pmatrix}
    \partial_q\lambda_+&\rbraket{+|\partial_q\halpha|-}\\
    \rbraket{-|\partial_q\halpha|+}&\partial_q\lambda_-
\end{pmatrix},
\end{align}\end{subequations}
and
\begin{subequations}\begin{align}
\tilde{W}&\equiv \tilde{A}^{-1}\partial_k^2M_k\tilde{A}=\diag(\tilde{w},\tilde{w}^*),\\
\tilde{w}&\equiv A^{-1}\partial_q^2\halpha A=\begin{pmatrix}
    \rbraket{+|\partial_q^2\halpha|+}&\rbraket{+|\partial_q^2\halpha|-}\\
    \rbraket{-|\partial_q^2\halpha|+}&\rbraket{-|\partial_q^2\halpha|-}
\end{pmatrix}.
\end{align}\end{subequations}
Here we used $\rbraket{\pm|\partial_q\alphaeff|\pm}=\partial_q\lambda_\pm$ as derived in the next subsection.
Then, we obtain
\begin{align}
    m_0''&=\Tr[\tilde{W}\tilde{P}_0]-\sum_{i\neq0}\frac{2}{m_i}\Tr[\tilde{V}\tilde{P}_i\tilde{V}\tilde{P}_0]\notag\\
    &=\Tr[\tilde{W}\tilde{P}_0]+\frac{1}{\Re[\lambda_-]}\Tr[\tilde{V}\tilde{P}_1\tilde{V}\tilde{P}_0]\notag\\
    &\qquad-2\Re\left[\frac{2}{\Delta\lambda}\Tr[\tilde{V}\tilde{P}_2\tilde{V}\tilde{P}_0]\right].\label{eq:m0pp}
\end{align}
To obtain the second line, we used the relations
\begin{align}
    \tau_x\tilde{V}^*\tau_x=-\tilde{V},\quad \tau_x\tilde{P}_3^*\tau_x=\tilde{P}_2,\quad \tau_x\tilde{P}_0^*\tau_x=\tilde{P}_0.
\end{align}

The first term of Eq.~\eqref{eq:m0pp} is given by
\begin{align}
    \Tr[\tilde{W}\tilde{P}_0]&=\Tr\left[\begin{pmatrix}\tilde{w}&0\\0&\tilde{w}^*\end{pmatrix}_\tau\frac{1-\tau_x}{2}\frac{1-\sigma_z}{2}\right]\notag\\
    &=\frac{1}{2}\tr\left[\tilde{w}\frac{1-\sigma_z}{2}\right]+\frac{1}{2}\tr\left[\tilde{w}^*\frac{1-\sigma_z}{2}\right]\notag\\
    &=\Re[\rbraket{-|\partial_q^2\halpha|-}]\notag\\
    &=\Re[\tr[\partial_q^2\halpha p_-]].
\end{align}
Here, ``$\tr$'' represents the trace in the $\sigma$ space.
The quantity in the last line can be rewritten as
\begin{align}
    \tr[\partial_q^2\halpha p_-]
    &=\partial_q\tr[\partial_q\halpha p_-]-\tr[\partial_q\halpha \partial_qp_-]\notag\\
    &=\partial_q^2\lambda_--\tr\left[\partial_q\halpha\frac{p_-\partial_q\halpha p_++p_+\partial_q\halpha p_-}{\lambda_--\lambda_+}\right]\notag\\
    &=\partial_q^2\lambda_-+\frac{2}{\Delta\lambda}\tr[\partial_q\halpha p_-\partial_q\halpha p_+].
\end{align}
The second line follows from the derivative of the projection operator discussed in the next subsection.

The second term of Eq.~\eqref{eq:m0pp} is given by
\begin{align}
    \frac{1}{\Re[\lambda_-]}\Tr[\tilde{V}\tilde{P}_1\tilde{V}\tilde{P}_0]&=\frac{T_1+T_{1C}}{\Re[\lambda_-]},
\end{align}
with
\begin{align}
    T_1&\equiv \Tr[\tilde{V}[\tilde{P}_1]_{C\to0}\tilde{V}\tilde{P}_0]\notag\\
    &=\Tr\left[\begin{pmatrix}
        \tilde{v}&0\\0&-\tilde{v}^*
    \end{pmatrix}\frac{1+\tau_x}{2}\frac{1-\sigma_z}{2}\right.\notag\\
    &\qquad\qquad\qquad\cdot\left.\begin{pmatrix}
        \tilde{v}&0\\0&-\tilde{v}^*
    \end{pmatrix}\frac{1-\tau_x}{2}\frac{1-\sigma_z}{2}\right]\notag\\
    &=\frac{1}{2}\Re\ \tr\left[\tilde{v}\frac{1-\sigma_z}{2}\tilde{v}\frac{1-\sigma_z}{2}+\tilde{v}\frac{1-\sigma_z}{2}\tilde{v}^*\frac{1-\sigma_z}{2}\right]\notag\\
    &=\frac{1}{2}\Re\left[(\partial_q\lambda_-)^2+\partial_q\lambda_-\partial_q\lambda_-^*\right]\notag\\
    &=(\partial_q\Re[\lambda_-])^2,
\end{align}
and
\begin{align}
    T_{1C}&=-\Tr\left[\begin{pmatrix}
        \tilde{v}&0\\0&-\tilde{v}^*
    \end{pmatrix}\sigma_-\begin{pmatrix}
        C&C^*\\C&C^*
    \end{pmatrix}_\tau\right.\notag\\
    &\qquad\qquad\qquad\left.\cdot\begin{pmatrix}
        \tilde{v}&0\\0&-\tilde{v}^*
    \end{pmatrix}\frac{1-\tau_x}{2}\frac{1-\sigma_z}{2}\right]\notag\\
    &=-\frac{1}{2}\left[C\tr\left[\tilde{v}\sigma_-\tilde{v}\frac{1-\sigma_z}{2}\right]+C^*\tr\left[\tilde{v}\sigma_-\tilde{v}^*\frac{1-\sigma_z}{2}\right]\right.\notag\\
    &+\left.C\tr\left[\tilde{v}^*\sigma_-\tilde{v}\frac{1-\sigma_z}{2}\right]+C^*\tr\left[\tilde{v}^*\sigma_-\tilde{v}^*\frac{1-\sigma_z}{2}\right]\right]\notag\\
    &=-2\Re\Bigl[C\partial_q\Re[\lambda_-]\rbraket{+|\partial_q\halpha|-}\Bigr].
\end{align}

The third term of Eq.~\eqref{eq:m0pp} is given by
\begin{align}
    -4\Re\left[\frac{T_2+T_{2C}}{\Delta\lambda}\right],
\end{align}
with
\begin{align}
    T_2&\equiv \Tr[\tilde{V}[\tilde{P}_2]_{C\to0}\tilde{V}\tilde{P}_0]\notag\\
    &=\Tr\left[\begin{pmatrix}
        \tilde{v}&0\\0&-\tilde{v}^*
    \end{pmatrix}\frac{1+\tau_z}{2}\frac{1+\sigma_z}{2}\right.\notag\\
    &\qquad\qquad\left.\begin{pmatrix}
        \tilde{v}&0\\0&-\tilde{v}^*
    \end{pmatrix}\frac{1-\tau_x}{2}\frac{1-\sigma_z}{2}\right]\notag\\
    &=\frac{1}{2}\tr\left[\tilde{v}\frac{1+\sigma_z}{2}\tilde{v}\frac{1-\sigma_z}{2}\right]\notag\\
    &=\frac{1}{2}\tr[\partial_q\halpha p_+\partial_q\halpha p_-],
\end{align}
and
\begin{align}
    T_{2C}&\equiv\Tr\left[\begin{pmatrix}
        \tilde{v}&0\\0&-\tilde{v}^*
    \end{pmatrix}\sigma_-\begin{pmatrix}
        C&0\\C&0
    \end{pmatrix}_\tau\right.\notag\\
    &\qquad\qquad\left.\cdot\begin{pmatrix}
        \tilde{v}&0\\0&-\tilde{v}^*
    \end{pmatrix}\frac{1-\tau_x}{2}\frac{1-\sigma_z}{2}\right]\notag\\
    &=\frac{C}{2}\Bigl[\tr\left[\tilde{v}\sigma_-\tilde{v}\frac{1-\sigma_z}{2}\right]+\tr\left[\tilde{v}^*\sigma_-\tilde{v}\frac{1-\sigma_z}{2}\right]\Bigr]\notag\\
    &=\frac{C}{2}\Bigl[\partial_q\lambda_-\rbraket{+|\partial_q\halpha|-}+\partial_q\lambda_-^*\rbraket{+|\partial_q\halpha|-}\Bigr]\notag\\
    &=C\partial_q\Re[\lambda_-]\rbraket{+|\partial_q\halpha|-}.
\end{align}
We obtain
\begin{align}
    &\frac{T_{1C}}{\Re[\lambda_-]}-4\Re\frac{T_{2C}}{\Delta\lambda}\notag\\
    &=-2\frac{1}{\Re[\lambda_-]}\Re[C\partial_q\Re[\lambda_-]\rbraket{+|\partial_q\halpha|-}]\notag\\
    &\qquad\qquad-4\Re\left[\frac{C\partial_q\Re[\lambda_-]\rbraket{+|\partial_q\halpha|-}}{\Delta\lambda}\right]\notag\\
    &=-2\partial_q\Re[\lambda_-]\Re\left[\frac{\rbraket{+|\partial_q\halpha|-}\left(\Delta\lambda+2\Re[\lambda_-]\right)}{\Re[\lambda_-]\Delta\lambda}\right.\notag\\
    &\qquad\qquad\qquad\qquad\cdot\left.\frac{-c\Re[\lambda_-]}{2\Re[\lambda_-]+\Delta\lambda}\right]\notag\\
    &=2\partial_q\Re[\lambda_-]\Re\left[\frac{c\rbraket{+|\partial_q\halpha|-}}{\Delta\lambda}\right].
\end{align}

Finally, we obtain
\begin{align}
    m_0''&=\partial_q^2\Re[\lambda_-]+\frac{(\partial_q\Re[\lambda_-])^2}{\Re[\lambda_-]}\notag\\
    &\quad+2\partial_q\Re[\lambda_-]\Re\left[\frac{\rbraket{+|\partial_q\halpha|-}\braket{-|+}}{\Delta\lambda}\right]\notag\\
    &=\frac{1}{-\Re[\lambda_-]}\partial_q[(-\Re[\lambda_-])\partial_q\Re[\lambda_-]]\notag\\
    &\quad+2\partial_q\Re[\lambda_-]\Re\left[\frac{\braket{-|\partial_q\halpha|-}-\partial_q\lambda_-}{\Delta\lambda}\right].
\end{align}
The first term has the same form as the local stability condition in equilibrium~\cite{Langer1967-rh,Samokhin2017-xf}, while the second term is negligible as we see below.

Let us estimate $m_0''$.
We obtain
\begin{align}
    -\Re[\lambda_-]&\simeq |\bar{\epsilon}(\mE)|-\xi_0^2q^2={|\bar{\epsilon}(\mE)|}(1-\xi^2q^2),
\end{align}
with $\xi\equiv \xi_0|\bar{\epsilon}(\mE)|^{-1/2}$.
We also have $\Delta\lambda\sim 2|a_x|$ and $\braket{-|\partial_q\halpha|-}\simeq 2\xi_0^2(q-q_\mE)$.
Let us write $Q=\xi q$ and $Q_\mE=\xi q_\mE$.
Then, we obtain
\begin{align}
    \frac{1}{\xi^{2}|\bar{\epsilon}(\mE)|}m_0''&=\frac{\partial_Q[(1-Q^2)2Q]}{1-Q^2}+4Q\frac{-2\xi_0^2q_\mE}{2|a_x|\xi}\notag\\
    &=\frac{\partial_Q[(1-Q^2)2Q]}{1-Q^2}-4Q\frac{|\bar{\epsilon}(\mE)|Q_\mE}{|a_x|}.\label{eq:dominant_m0pp}
\end{align}
In the absence of the second term, the stability condition $m_0''>0$ is equivalent to $|Q|<Q_{\rm c}=1/\sqrt{3}$ as in equilibrium~\cite{Langer1967-rh,Samokhin2017-xf}.
This remains valid even in the presence of $\mE$ since the second term makes only a small correction to $Q_{\rm c}$ of the order $\sqrt{{|\bar{\epsilon}(\mE)|}}\xi_0q_\mE/|a_x|\ll1$ and is negligible.
Thus, the steady states satisfying $|\xi q|<1/\sqrt{3}$ are stable against perturbations.

\subsection{Derivation of Eq.~\eqref{eq:formula_Hel}}
Let us derive Eq.~\eqref{eq:formula_Hel}.
We consider an eigenvalue $m_i(k)$ of the matrix $M_k=\sum_im_i(k)P_i(k)$.
We can write
\begin{align}
    m_i(k)&=\frac{\Tr[M_kP_i(k)]}{\Tr[P_i(k)]}.
\end{align}
We assume that $\Tr[P_i(k)]$, namely the degree of degeneracy, does not change for sufficiently small changes of $k$.
By using the formula for the projection operator~\cite{Ashida2020-yn}
\begin{align}
    P_i(k)&=\oint_{C_i}\frac{dz}{2\pi i}\frac{1}{z-M_k},
\end{align}
with $C_i$ an anticlockwise tiny loop around $z=m_i(k)$,
we obtain
\begin{align}
    \Tr[P_i(k)]\partial_km_i(k)&=\Tr[\partial_kM_k\,P_i(k)],
\end{align}
by using $\oint dz(z-M_k)^{-2}=0$.
In particular, this means $\rbraket{\pm|\partial_q\alphaeff|\pm}=\partial_q\lambda_\pm$ for the case of $\alphaeff$.
We also obtain
\begin{align}
&\Tr[P_i(k)]\partial_k^2m_i(k)\notag\\
&=\Tr[\partial_k^2M_kP_i(k)]\notag\\
&\qquad +\Tr\left[\partial_kM_k\oint_{C_i}\frac{dz}{2\pi i}\frac{1}{z-M_k}\partial_kM_k\frac{1}{z-M_k}\right]\notag\\
&=\Tr[\partial_k^2M_kP_i(k)]+\sum_{ab}\oint_{C_i}\frac{dz}{2\pi i}\frac{J_{ab}(k)}{[z-m_a(k)][z-m_b(k)]}\notag\\
&=\Tr[\partial_k^2M_kP_i(k)]+\sum_{a\neq i}\frac{J_{ai}(k)+J_{ia}(k)}{m_i(k)-m_a(k)},
\end{align}
with 
\begin{align}
    J_{ab}(k)&=\Tr[\partial_kM_k P_a(k)\partial_kM_kP_b(k)]=J_{ba}(k).
\end{align}
This gives Eq.~\eqref{eq:formula_Hel} by taking $k\to0$ and $\Tr[P_0(k)]=1$.

\section{unidirectional superconductivity in general bilayer GL model}
In the remaining part of the Supplemental Material, we discuss the generalization of our model to 
the general GL model of bilayer superconductors.
This ensures that our model is not oversimplified.
\subsection{Setup}
We start with the GL free energy of the form
\begin{align}
    F[\bm{\psi}]&=N_0\int dx\,\bm{\psi}^\dagger(x)\halpha(\nabla/i)\bm{\psi}(\bm{x})\notag\\
    &\qquad+\frac{1}{2}\beta_{ijkl}\psi_i^*(x)\psi_j^*(x)\psi_k(x)\psi_l(x).
\end{align}
Here, the GL coefficient $\beta_{ijkl}$ satisfies
\begin{align}
    \beta_{ijkl}=\beta_{jikl}=\beta_{ijlk},\quad \beta_{ijkl}^*=\beta_{lkji}.
\end{align}
It is sufficient to 
neglect $O(\psi^6)$ terms and gradient terms in the quartic term by assuming a second-order transition in equilibrium, since they are of higher order in the reduced temperature: $\nabla/i\sim \xi^{-1}\sim\sqrt{\epsilon}$ and $\psi\sim \sqrt{\epsilon}$.
The TDGL equation for the plain-wave ansatz is given by
\begin{align}
    [\partial_t+i\Phi]\bm{\psi}(q,t)=-[\halpha(q)+\hbeta(\bm{\psi}(q,t))]\bm{\psi}(q,t),
    \label{eq:TDGL_momentum_SM}
\end{align}
with
\begin{align}
    [\hbeta(\bm{\psi}(q,t))]_{ij}\equiv \beta_{iabj}\psi_a^*(q,t)\psi_b(q,t).
\end{align}
By using $\bm{\psi}(q,t)=e^{i\chi(q,t)}R(q,t)\hat{\psi}(q,t)$, the TDGL equation is rewritten as
\begin{subequations}\begin{align}
\!\!\!\!\!\!\partial_t\chi(q,t)&=-\bkpsit{\Phi}+i\hat{\psi}(q,t)^\dagger\partial_t\hat{\psi}(q,t),\\
\!\!\!\!\!\!\partial_tR(q,t)&=-\big(\bkpsit{\hat{\alpha}(q)}\notag\\
\!\!\!\!\!\!&\quad+ R^2(q,t)\braket{\hbeta(\hat{\psi}(q,t))}_{\hat{\psi}(q,t)}\big)R(q,t),\\
\!\!\!\!\!\!(1-\hat{\psi}(q,t)&\hat{\psi}(q,t)^\dagger)\partial_t\hat{\psi}(q,t)\\
\!\!\!\!\!\!&=-\left[\alphaeff(q)-\braket{\alphaeff(q)}_{\hat{\psi}(q,t)}+R^2(q,t)\right.\notag\\
\!\!\!\!\!\!&\left.\cdot\left\{\hbeta(\psi(q,t))-\braket{\hbeta(\psi(q,t))}_{\hat{\psi}(q,t)}\right\}\right]\hat{\psi}(q,t).\notag
\end{align}\end{subequations}
Thus, the steady state is given by the solution of
\begin{subequations}\begin{align}
    \partial_t\chi_\sst(q,t)&=-\braket{\Phi}_{\hat{\psi}_\sst(q)},\\
    R^2_\sst(q)&=-\frac{\braket{\halpha(q)}_{\hat{\psi}_\sst(q)}}{\braket{\beta(\hat{\psi}_\sst(q))}_{\hat{\psi}_\sst(q)}},\label{eq:R2ss}
\end{align}\end{subequations}
    as well as
\begin{align}
&\Bigl[\alphaeff(q)-\braket{\alphaeff(q)}_{\hat{\psi}_\sst(q)}]\hat{\psi}_\sst(q)\label{eq:TDGL_internal_SM}\\
&=\frac{\braket{\halpha(q)}_{\hat{\psi}_\sst(q)}}{\braket{\beta(\hat{\psi}_\sst(q))}_{\hat{\psi}_\sst(q)}}\left[\hbeta(\hat{\psi}_\sst(q))-\braket{\hbeta(\hat{\psi}_\sst(q))}_{\hat{\psi}_\sst(q)}\right]\hat{\psi}_\sst(q).\notag
\end{align}

\subsection{Steady states}
Equation~\eqref{eq:TDGL_internal_SM} implies the deviation of $\hat{\psi}_\sst(q)$ from the eigenstates of $\alphaeff$ is of the order $\braket{\halpha(q)}\sim|\epsilon|$.
To see this, let us focus on Eq.~\eqref{eq:R2ss}.
It is natural to assume the denominator to be positive for any states we are interested in, since otherwise we need a $O(\psi^6)$ term in the GL free energy even in equilibrium.
Thus, we can limit ourselves to the situation where
\begin{align}
\braket{\hat{\alpha}(q)}_{\hat{\psi}_\sst(q)}=\Re[\braket{\alphaeff(q)}_{\hat{\psi}_\sst(q)}]
\end{align}
is negative, to have a finite amplitude $R_\sst(q)>0$.
By writing the eigenvalues of $\halpha(q)$ as $\alpha_\pm(q)$ with $\alpha_+(q)\ge\alpha_-(q)$, the lefthand side is larger than or equal to $\alpha_-(q)$.
Thus, steady states are possible only when $\alpha_-(q)<0$, i.e., when the system is superconducting in the absence of $\Phi$.
There is a minimum of $\alpha_-(q)$ and let us assume it is achieved at $q=q_0$, where $q_0$ vanishes in the presence of either inversion or time-reversal symmetry.
We obtain
\begin{align}
    \alpha_-(q)=\epsilon+\xi_0^2(q-q_0)^2+O(q-q_0)^3,
\end{align}
where we introduced $\epsilon=\alpha_-(q_0)$, $\xi_0^2=[\partial_{q}^2\alpha_-(q)]_{q=q_0}/2$, and used $[\partial_q\alpha_-(q)]_{q=q_0}=0.$
We obtain $\epsilon<0$ and $\xi_0^2>0$ by assumption.
The equilibrium momentum $q_0$ can be erased by 
rewriting $q-q_0$ as $q$, and thus we set $q_0=0$ in the following.
In summary, we know that $\alpha_-(q)$ and $q$ are of the order $|\epsilon|$ and $|\epsilon|^{1/2}$, and therefore $R^2_{\sst}(q)=O(|\epsilon|)$.
Since the other terms in the right-hand side of Eq.~\eqref{eq:TDGL_internal_SM} are not singular in terms of $\epsilon$, the right-hand side of Eq.~\eqref{eq:TDGL_internal_SM} is $O(|\epsilon|)$.

Let us explicitly construct a perturbative solution of Eq.~\eqref{eq:TDGL_internal_SM}.
We abbreviate the argument $q$ of quantities for a while.
When we substitute the right eigenstate $\ket{-}$ of $\alphaeff$ for $\hat{\psi}_\sst$, the left-hand side vanishes while the right-hand side is $O(|\epsilon|)$, and thus $\ket{-}$ is a solution when $\epsilon\to0$.
Let us write
\begin{align}
    \hat{\psi}_\sst=\varphi_+\ket{+}+\varphi_-\ket{-},
\end{align}
with $|\varphi_+|\ll1$, $\varphi_-=1+\delta\varphi_-\in\mathbb{R}$ and $|\delta\varphi_-|\ll1$.
The normalization condition reads
\begin{align}
    1&=\hat{\psi}_\sst^\dagger\hat{\psi}_\sst
    =|\varphi_-|^2+|\varphi_+|^2+c(\varphi_-^*\varphi_++\varphi_+^*\varphi_-).
\end{align}
Here, we choose the eigenstates to satisfy $c=\braket{-|+}\in\mathbb{R}$.
Then, we can explicitly evaluate
\begin{align}
    \braket{\alphaeff}_{\hat{\psi}_\sst}
    &=\hat{\psi}_\sst^\dagger(\lambda_-\varphi_-\ket{-}+\lambda_+\varphi_+\ket{+})\notag\\
    &=\hat{\psi}_\sst^\dagger(\lambda_-\hat{\psi}_\sst+\Delta\lambda\varphi_+\ket{+})\notag\\
    &=\lambda_-+\Delta\lambda\varphi_+(\varphi_+^*+\varphi_-^*c).\label{eq:alphaeff_temp112}
\end{align}
We obtain
\begin{align}
    &[\alphaeff-\braket{\alphaeff}_{\hat{\psi}_\sst}]\hat{\psi}_\sst\notag\\
    &=\Delta\lambda\varphi_+\ket{+}-\Delta\lambda\varphi_+(\varphi_+^*+\varphi_-^*c)[\varphi_+\ket{+}+\varphi_-\ket{-}]\notag\\
    &=\Delta\lambda\varphi_+\Bigl[\bigl\{1-\varphi_+(\varphi_+^*+\varphi_-^*c)\bigr\}\ket{+}\notag\\
    &\qquad\qquad\qquad-\varphi_-(\varphi_+^*+\varphi_-^*c)\ket{-}\Bigr]\notag\\
    &\sim \Delta\lambda\varphi_+\Bigl[\ket{+}-c\ket{-}\Bigr],\label{eq:107LHS}
\end{align}
to the leading order of $\varphi_+$ and $\delta\varphi_-$, which are $o(1)$ in terms of $\epsilon$.
We also obtain
\begin{align}
    &\frac{\braket{\halpha}_{\hat{\psi}_\sst}}{\braket{\beta(\hat{\psi}_\sst)}_{\hat{\psi}_\sst}}\left[\hbeta(\hat{\psi}_\sst)-\braket{\hbeta(\hat{\psi}_\sst)}_{\hat{\psi}_\sst}\right]\hat{\psi}_\sst\notag\\
    &\simeq \frac{\braket{\halpha}_{\hat{\psi}_\sst}}{\beta_{----}}[\hat{\beta}(\ket{-})\ket{-}-\ket{-}\braket{-|\hat{\beta}(\ket{-})|-}]\notag\\
    &=\frac{\braket{\halpha}_{\hat{\psi}_\sst}}{\beta_{----}}[1-\ket{-}\bra{-}]\hat{\beta}(\ket{-})\ket{-}\label{eq:RHStemp}\\
    &=\frac{\braket{\halpha}_{\hat{\psi}_\sst}}{\beta_{----}}[\ket{+}-c\ket{-}]\rbraket{+|\hat{\beta}(\ket{-})|-},\notag
\end{align}
to the leading order, by using $\braket{\halpha}_{\hat{\psi}_\sst}=O(|\epsilon|)$.
We also used $1-\ket{-}\bra{-}=(1-\ket{-}\bra{-})(p_++p_-)=(\ket{+}-c\ket{-})\rbra{+}$ and $\beta_{----}\equiv\bra{-}_i\bra{-}_j\beta_{ijkl}\ket{-}_k\ket{-}_l$.
Comparing Eqs.~\eqref{eq:107LHS} and~\eqref{eq:RHStemp}, $\varphi_+$ must be $O(|\epsilon|)$ since $\rbra{+}\hat{\beta}(\ket{-})\ket{-}$ is not singular in terms of $\epsilon$.
According to $\varphi_+=O(|\epsilon|)$, we also know that
\begin{align}
\Re[\lambda_-]\simeq
{\braket{\halpha}_{\hat{\psi}_\sst}}-\Re[\Delta\lambda\varphi_+c].
\end{align}
must be $O(|\epsilon|)$.

The lowest-order term of $\varphi_+$ is explicitly obtained as follows.
By using the relation $\braket{\halpha}_{\hat{\psi}_\sst}=\Re[\braket{\alphaeff}_{\hat{\psi}_\sst}$] and Eqs.~\eqref{eq:alphaeff_temp112} and~\eqref{eq:RHStemp}, 
we obtain to the leading order
\begin{subequations}
    \begin{align}
    \Re[\Delta\lambda\varphi_+]&=\frac{\Re[\lambda_-]+c\Re[\Delta\lambda\varphi_+]}{\beta_{----}}\Re[\beta'_{+---}],\\
    \Im[\Delta\lambda\varphi_+]&=\frac{\Re[\lambda_-]+c\Re[\Delta\lambda\varphi_+]}{\beta_{----}}\Im[\beta'_{+---}],
\end{align}
\end{subequations}
with $\beta'_{+---}\equiv\rbraket{+|\hat{\beta}(\ket{-})|-}$.
This explicitly confirms that $\varphi_+$ has an $O(\epsilon)$ solution.
We obtain $\delta\varphi_-=\varphi_--1$ from the normalization condition
\begin{align}
    0&\simeq 2\delta\varphi_-+2c\Re[\varphi_+],
\end{align}
i.e., 
\begin{align}
\varphi_-\simeq 1-c\Re[\varphi_+].
\end{align}

\subsection{Physical properties}
Let us evaluate the critical electric field based on the above results.
We recover the argument $q$ of quantities in the following for a while.
In the basis diagonalizing $\halpha(q)$, $\alphaeff(q)$ is given by
\begin{align}
\alpha_-(q)+\begin{pmatrix}
    \Delta\alpha(q)+i\Phi_{++}(q)&i\Phi_{+-}(q)\\
    i\Phi_{-+}(q)&i\Phi_{--}(q)
\end{pmatrix},
\end{align}
where $\Phi_{ss'}(q)\equiv\hat{\psi}^\dagger_s(q)\Phi\hat{\psi}_{s'}(q)$ with $s,s'=\pm$ and $\halpha_s(q)\hat{\psi}_s(q)=\alpha_s(q)\hat{\psi}_s(q)$.
Here, $\Delta\alpha(q)\gg|\alpha_-(q)|=O(|\epsilon|)$ is satisfied.
Let us assume $\Delta\alpha(q)$ is also larger than $\mE$ and thus $|\Phi_{ss'}|$.
The consistency of the assumption is validated below.
The eigenvalues $\lambda_\pm(q)$ of $\alphaeff(q)$ are given by
\begin{align}
\lambda_\pm(q)
&\simeq \alpha_\pm(q)+i\Phi_{\pm\pm}(q) \mp\frac{|\Phi_{+-}(q)|^2}{\Delta\alpha(q)}.
\end{align}
For $\Re[\lambda_-(q)]$ to be $O(|\epsilon|)$, $\mE$ must be $O(|\epsilon|^{1/2})$.
It also follows that $c=\braket{-|+}$ is also $O(|\epsilon|^{1/2})$, and thus we obtain
\begin{align}
R^2_\sst(q)&\propto\braket{\halpha(q)}_{\hat{\psi}_\sst(q)}\notag\\
&\simeq\Re[\lambda_-(q)]\\
&\simeq \Re[\alpha_-(q)]+\frac{|\Phi_{+-}(q)|^2}{\Delta\alpha(q)}.
\end{align}
The critical electric field is determined so that the right-hand side vanishes for $q=0$.
Note that 
\begin{align}
|\Phi_{+-}|^2
&=\tr[p_{0+}\Phi p_{0-}\Phi]\notag\\
&=\tr[p_{0+}\Phi[p_{0-},\Phi]]\notag\\
&=-\Re\,\tr[p_{0+}\Phi[p_{0+},\Phi]]\notag\\
&=-\frac{1}{2}\Re\,\tr[[p_{0+},\Phi]^2]\notag\\
&=-\frac{1}{2\Delta\alpha^2}\tr[[\halpha,\Phi]^2]\notag\\
&=-\frac{1}{2\Delta\alpha^2}\mE^2\tr[(2i\alpha_x\sigma_y-2i\alpha_y\sigma_x)^2]\notag\\
&=\frac{\alpha_x^2+\alpha_y^2}{(\Delta\alpha/2)^2}\mE^2,
\end{align}
where we abbreviated the argument $q$ and used $p_{0+}=[\halpha-\alpha_-]/\Delta\alpha$
with $p_{0s}\equiv\hat{\psi}_{s}\hat{\psi}_{s}^\dagger$.
By using $\Delta\alpha(q)/2=\sqrt{\alpha_x(q)^2+\alpha_y(q)^2+\alpha_z(q)^2}$, we obtain the critical electric field
\begin{align}
\mE^2_{\rm c}(\epsilon)&=[-\epsilon]\frac{2[\Delta\alpha(0)/2]^3}{\alpha_x(0)^2+\alpha_y(0)^2}.
\end{align}
This is consistent with the result in the main text.
Since the critical electric field is $O(|\epsilon|^{1/2})$, the validity of the assumption $\Delta\alpha(q)\gg |\mE|$ has been confirmed.

The electric current is given by
\begin{align}
&R_\sst^2\hat{\psi}_\sst^\dagger\partial_q\halpha\hat{\psi}_\sst\notag\\
&\simeq -\frac{\Re[\lambda_-]+c\Re[\Delta\lambda\varphi_+]}{\beta_{----}}(\braket{-|\partial_q\halpha|-}+O(\epsilon))\notag\\
&=\frac{-\Re[\lambda_-]}{\beta_{----}}(1+O(|\epsilon|^{1/2}))(\braket{-|\partial_q\halpha|-}+O(\epsilon)).
\end{align}
The quantity $\braket{-|\partial_q\halpha|-}$ contributes by $O(q)=O(|\epsilon|^{1/2})$, and we obtain the leading-order contribution
\begin{align}
&R_\sst^2\hat{\psi}_\sst^\dagger\partial_q\halpha\hat{\psi}_\sst 
=\frac{-\Re[\lambda_-]}{\beta_{----}}\braket{-|\partial_q\halpha|-},
\end{align}
Thus, neglecting the $O(|\epsilon|^{1/2})$ correction, we arrive at the same expression of the electric current as that for the minimal model by defining
\begin{align}
\beta_0\equiv\beta_{----}=\bra{-}_i\bra{-}_j\beta_{ijkl}\ket{-}_k\ket{-}_l.
\end{align}
This means that a parallel discussion leads to the unidirectional superconductivity.

Let us express the relevant quantities
in terms of the matrix elements of $\halpha(q)$.
We can write
\begin{align}
-\Re[\lambda_-(q)]&\simeq -\epsilon-\xi_0^2q^2-\frac{2|\Phi_{+-}(0)|^2}{\Delta\alpha(0)}\notag\\
&=|\bar{\epsilon}(\mE)|-\xi_0^2q^2.
\end{align}
We also obtain
\begin{align}
\braket{-|\partial_q\halpha|-}
&=\Re[\partial_q\lambda_-]+\braket{-|+}\rbraket{+|\partial_q\halpha|-}\notag\\
&=2\xi_0^2q+[\braket{-|+}\rbraket{+|\partial_q\halpha|-}]_{q=0}+O(\epsilon),\notag\\
&\equiv 2\xi_0^2(q-q_\mE)+O(\epsilon),
\end{align}
with $2\xi_0^2q_\mE\equiv -[\braket{-|+}\rbraket{+|\partial_q\halpha|-}]_{q=0}$,
since $\braket{-|+}\sim O(|\epsilon|^{1/2})$.
It is sufficient to obtain $q_\mE$ up to $O(\mE)$.
We obtain
\begin{align}
&-2\xi_0^2q_\mE\notag\\
&=\Re\left[\frac{\tr[p_-^\dagger p_+\partial_q\halpha p_-]}{\tr[p_-^\dagger p_-]}\right]\notag\\
&=\Re\left[\frac{\tr[[p_-,p_-^\dagger] p_+\partial_q\halpha ]}{\tr[p_-^\dagger p_-]}\right]\notag\\
&\simeq\Re\left[\frac{2i}{|\Delta\alpha|^2}\Tr[[\Phi,\halpha]p_{0+}\partial_q\halpha]\right]\notag\\
&=-\Im\left[\frac{2}{\Delta\alpha^3}\tr[[\Phi,\halpha-\alpha_0]\left(\halpha-\alpha_0+\frac{\Delta\alpha}{2}\right)\partial_q(\halpha-\alpha_0)]\right]\notag\\
&=-\Im\left[\frac{1}{\Delta\alpha^3}\tr[[\Phi,\halpha-\alpha_0][\halpha-\alpha_0,\partial_q(\halpha-\alpha_0)]]\right]\notag\\
&\qquad-\Im\left[\frac{1}{\Delta\alpha^2}\tr[[\Phi,\halpha-\alpha_0]\partial_q(\halpha-\alpha_0)]\right]\notag\\
&=-\Im\left[\frac{1}{\Delta\alpha^3}\tr[[\Phi,\halpha][\halpha,\partial_q\halpha]]\right]-\Im\left[\frac{1}{\Delta\alpha^2}\tr[[\Phi,\halpha]\partial_q\halpha]\right]\notag\\
&=\frac{i}{\Delta\alpha^2}\tr[[\halpha,\partial_q\halpha]\Phi]\notag\\
&=\frac{i}{\Delta\alpha^2}4i(-\mE)\hat{z}\cdot\bm{\alpha}\times\partial_q\bm{\alpha}\notag\\
&=\frac{\alpha_x\partial_q\alpha_y-\alpha_y\partial_q\alpha_x}{(\Delta\alpha/2)^2}\mE.
\end{align}
Here we used
\begin{align}
p_\pm&=\frac{1}{2}\left(1\pm\frac{\alphaeff-(\lambda_++\lambda_-)/2}{\Delta\lambda/2}\right).
\end{align}
We also used the fact that $\partial_q(\halpha-\alpha_0)^2=\partial_q(\Delta\alpha/2)^2$ is proportional to the identity operator.
The second term in the numerator of the last line can be finite in the absence of time-reversal symmetry.
The conversion efficiency $\mC$ is given by 
\begin{align}
\mC&=\xi_0\frac{dq_\mE}{d\mE}\mEc(-1)\notag\\
&=\xi_0\frac{1}{-2\xi_0^2}\frac{\alpha_x\partial_q\alpha_y-\alpha_y\partial_q\alpha_x}{[\Delta\alpha/2]^2}\left(\frac{2[\Delta\alpha/2]^{3}}{\alpha_x^2+\alpha_y^2}\right)^{1/2}\notag\\
&=-\frac{1}{\sqrt{2}}\frac{\xi_0^{-1}[\alpha_x\partial_q\alpha_y-\alpha_y\partial_q\alpha_x]}{(\alpha_x^2+\alpha_y^2+\alpha_z^2)^{1/4}\sqrt{\alpha_x^2+\alpha_y^2}}.
\label{eq:C_general}
\end{align}
Equation~\eqref{eq:C_general} is reduced to the formula for $\mC$ in the main text for the minimal model.

\section{Local stability of the general bilayer TDGL model with and without current bias}
\subsection{Local stability without the current bias}
We first discuss the local stability of the steady states for the general bilayer TDGL model without the current bias.
By considering the perturbed state of the form $\bm{\psi}'(x,t)=e^{i\chi_\sst(t)+iqx}[\bm{\psi}_\sst(q)+\delta\bm{\psi}(x,t)]$,
the stability of the steady state is described by the eigenvalues of the matrix
\begin{widetext}
\begin{align}
M_k(q)&=\begin{pmatrix}
    \alphaeff(q+k)-i\braket{\Phi}_{\hat{\psi}_\sst}+2\hbeta(\bm{\psi}_\sst)&\tilde{\beta}(\bm{\psi}_\sst)\\\tilde{\beta}(\bm{\psi}_\sst)^*&[\alphaeff(q-k)-i\braket{\Phi}_{\hat{\psi}_\sst}+2\hbeta(\bm{\psi}_\sst)]^*
\end{pmatrix},\label{eq:Mkq_general}
\end{align}    
\end{widetext}
with $[\tilde{\beta}(\bm{\psi})]_{ij}\equiv\beta_{ijab}{\psi}_a\psi_b$.
In the same way as before, the matrix $M_k(q)$ satisfies $\tau_xM_k(q)^*\tau_x=M_{-k}(q)$ and has a zero mode at $k=0$.
Thus, we are interested in the sign of $m_0''$ given by Eq.~\eqref{eq:formula_Hel}. 
The dominant contribution to $m_0''$ is $O(1)$ with respect to $\epsilon$ as seen in Eq.~\eqref{eq:dominant_m0pp}.
Below, we show that $m_0''$ coincides with $\Re[\lambda_-]^{-1}\partial_q(\Re[\lambda_-]\partial_q\Re[\lambda_-])$ except for $o(1)$ corrections in terms of $\epsilon$.
Here and hereafter, the argument $q$ is abbreviated unless necessary.

Let us first focus on $M_0$.
According to the TDGL equation, the steady state is the solution of
\begin{align}
[\alphaeff+\beta(\bm{\psi}_\sst)]\bm{\psi}_\sst=i\braket{\Phi}_{\hat{\psi}_\sst}\bm{\psi}_\sst.
\end{align}
We define
\begin{align}
\hgamma\equiv \alphaeff+\hbeta(\bm{\psi}_\sst),
\end{align}
and
\begin{align}
\hgamma\ket{\pm}=\gamma_\pm\ket{\pm},\quad \gamma_-=i\braket{\Phi}_{\ket{-}},\quad \ket{-}=\hat{\psi}_\sst.
\end{align}
In this section, $\ket{\pm}$ represents eigenstates of $\hgamma$ rather than those of $\alphaeff$, while they coincide with each other when the $O(\epsilon)$ correction is neglected.
We can write
\begin{align}
M_0&=\begin{pmatrix}
    \hgamma-\gamma_-&0\\
    0&[\hgamma-\gamma_-]^*
\end{pmatrix}\notag\\
&\qquad+|\bm{\psi}_\sst|^2\begin{pmatrix}
    \hbeta(\ket{-})&\tilde{\beta}(\ket{-})\\
    \tilde{\beta}(\ket{-})^*&\hbeta(\ket{-})^*
\end{pmatrix}.\label{eq:M0_def}
\end{align}
The first term of Eq.~\eqref{eq:M0_def} has a spectral decomposition
\begin{align}
\begin{pmatrix}
    \hgamma-\gamma_-&0\\
    0&[\hgamma-\gamma_-]^*
\end{pmatrix}=\Delta\gamma P'_2+\Delta\gamma^*P'_3,
\end{align}
with $\Delta\gamma\equiv\gamma_+-\gamma_-$ and
\begin{align}
P'_2=\frac{1+\tau_z}{2}p_+,\quad P'_3=\frac{1-\tau_z}{2}p_+^*,
\end{align}
as well as $p_\pm\equiv \ket{\pm}\rbra{\pm}$.
We also define
\begin{align}
\mP=\begin{pmatrix}
p_-&0\\0&p_-^*
\end{pmatrix},\quad \mQ=P'_2+P'_3=\begin{pmatrix}
p_+&0\\0&p_+^*
\end{pmatrix}.
\end{align}
Then, we obtain the decomposition
\begin{align}
M_0=\mM_0+\delta\mM,
\end{align}
with
\begin{subequations}\begin{align}
\mM_0&\equiv \mP M_0\mP+\mQ M_0\mQ,\\
\delta\mM&\equiv\mP M_0\mQ+\mQ M_0\mP\notag\\
&=O(|\bm{\psi}_\sst|^2)=O(|\epsilon|).
\end{align}\end{subequations}

The most important part is $\mP M_0\mP$ and is given by
\begin{align}
\mP M_0\mP&=\beta_-|\bm{\psi}_\sst|^2\begin{pmatrix}
    p_-&\ket{-}\rbra{-}^*\\
    \ket{-}^*\rbra{-}&p_-^*
\end{pmatrix},
\end{align}
with $\beta_-\equiv\beta_{ijkl}\rbra{-}_i\bra{-}_j\ket{-}_k\ket{-}_l$.
We thus define
\begin{subequations}\begin{align}
P_0&=\frac{1}{2}\begin{pmatrix}
    p_-&-\ket{-}\rbra{-}^*\\
    -\ket{-}^*\rbra{-}&p_-^*
\end{pmatrix},\\
P_1&=\frac{1}{2}\begin{pmatrix}
    p_-&\ket{-}\rbra{-}^*\\
    \ket{-}^*\rbra{-}&p_-^*
\end{pmatrix},
\end{align}\end{subequations}
which satisfy $\mP=P_0+P_1$ and $\mP\mM_0\mP=2\beta_-|\bm{\psi}_\sst|^2P_1$.
We also define the spectral decomposition
\begin{align}
\mQ M_0\mQ=m_2P_2+m_3P_3,
\end{align}
where $m_2=m_3^*$ has the magnitude $|\Delta\gamma|+O(\epsilon)$, with $P_3=\tau_xP_2\tau_x^*$ and $P_2+P_3=\mQ$.
Thus, we can write
\begin{align}
\mM_0=m_0P_0+m_1P_1+m_2P_2+m_3P_3,
\end{align}
with $m_0=0$ and $m_1=2\beta_-|\bm{\psi}_\sst|^2$.
Here, $P_0$ is also the exact projection operator of the zero mode of $M_0$, while $P_{1,2,3}$ are not due to $\delta\mM$.

We consider a non-Hermitian analog of the Schrieffer-Wolff transformation 
\begin{align}
e^{S}M_0e^{-S}=\mM_0+O(\epsilon^2/|\Delta\gamma|),
\end{align}
to trace out $\delta\mM$,
where
\begin{align}
S=\sum_{i=\{0,1\},j=\{2,3\}}\frac{P_i\delta\mM P_j-P_j\delta\mM P_i}{m_i-m_j},
\end{align}
is an $O(\epsilon/|\Delta\gamma|)$ matrix.
Thus, we obtain the spectral decomposition of $M_0$,
\begin{align}
M_0=\sum_im_ie^{-S}P_ie^{S}+O(\epsilon^2/|\Delta\gamma|).\label{eq:SD_M0}
\end{align}

According to Eq.~\eqref{eq:formula_Hel}, we obtain
\begin{align}
m_0''&\simeq\Tr[W e^{-S}P_0e^S]\notag\\
&\quad+2\sum_{i\neq 0}\frac{\Tr[Ve^{-S}P_0e^SVe^{-S}P_ie^S]}{m_0-m_i}\notag\\
&\simeq\bar{m}_0''+\delta m_0''.
\end{align}
Here we defined
\begin{align}
V=\diag(\partial_q\halpha,-\partial_q\halpha^*),\quad W=\diag(\partial_q^2\halpha,\partial_q^2\halpha^*),
\end{align}
and
\begin{subequations}\begin{align}
\bar{m}_0''&\equiv\Tr[W P_0]+2\sum_{i\neq 0}\frac{\Tr[VP_0VP_i]}{m_0-m_i},\\
\delta m_0''&\equiv\Tr[[S,W]P_0]\\
&\quad+2\sum_{i\neq0}\frac{\Tr[[S,V]P_0VP_i]+\Tr[VP_0[S,V]P_i]}{m_0-m_i}.\notag
\end{align}\end{subequations}
Clearly, $\bar{m}_0''$ is the main contribution of $O(1)$ as illustrated for the case of the minimal model.
While most terms of $\delta m_0''$ are $O(\epsilon)$ and can be neglected [note that matrix elements of $W$ and $V$ are $O(1)$], the second term with $i=1$ seems nontrivial at a glance since $S\sim m_0-m_1=O(\epsilon)$.
However, the $i=1$ contribution includes terms such as
\begin{align}
VP_1VP_0&=V'\tau_zP_1\tau_zV'P_0\notag\\
&=V'P_0V'P_0\notag\\
&=V'P_0\Re[\rbraket{-|\partial_q\halpha|-}],
\end{align}
and $VP_0VP_1=V'P_1\Re[\rbraket{-|\partial_q\halpha|-}]$,
by using $V'=\diag(\partial_q\halpha,\partial_q\halpha^*)$,
and hence is $O(|\epsilon|^{1/2})$.
Thus, $\delta m_0''$ is negligible and we discuss $\bar{m}_0''$ below.

In the following, we can identify $\ket{\pm}$ are the eigenstates of $\alphaeff$ to obtain the leading-order contribution.
By writing $P_0=\ket{\psi_-}\rbra{\psi_-}$ and
\begin{align}
 \!\!\!\!\ket{\psi_-}\equiv\frac{1}{\sqrt{2}}\begin{pmatrix}
    \ket{-}\\-\ket{-}^*
\end{pmatrix},\ \rbra{\psi_-}\equiv\frac{1}{\sqrt{2}}\left(\rbra{-},\,-\rbra{-}^*\right),
\end{align}
the first term of $\bar{m}_0''$ is given by
\begin{align}
\Tr[WP_0]&=\rbraket{\psi_-|W|\psi_-}\\
&=\Re[\rbraket{-|\partial_q^2\halpha|-}]\notag\\
&\simeq \partial_q\Re[\lambda_-]+\Re\left[\frac{2}{\Delta\lambda}\tr[\partial_q\halpha p_-\partial_q\halpha p_+]\right].\notag
\end{align}
The $i=1$ contribution from the second term of $\bar{m}_0''$ is
\begin{align}
\frac{2}{m_0-m_1}\Tr[VP_0VP_1]
&=-\frac{2}{m_1}\Re[\rbraket{-|\partial_q\halpha|-}]^2\notag\\
&=-\frac{1}{\beta_-|\bm{\psi}_\sst|^2}\Re[\partial_q\lambda_-]^2\notag\\
&\simeq \frac{\Re[\partial_q\lambda_-]^2}{\Re[\lambda_-]},
\end{align}
by using $\beta_-=\beta_0(1+O(c))\simeq \beta_0$.

The $i=2,3$ contributions are given as follows.
Note that
\begin{align}
\sum_{i=2,3}\frac{P_i}{m_0-m_i}&=-\frac{m_2^*P_2+m_2P_3}{|m_2|^2}\notag\\
&=-\frac{\Re[m_2][P_2+P_3]-i\Im[m_2][P_2-P_3]}{|m_2|^2}\notag\\
&=-\frac{2\Re[m_2]\mQ-\mQ M_0\mQ}{|m_2|^2},
\end{align}
as obtained from $\mQ M_0\mQ=m_2P_2+m_3P_3$ and $P_2+P_3=\mQ$.
This is approximated by that in the absence of $\delta\mM=O(\epsilon)$, and thus
\begin{align}
\sum_{i=2,3}\frac{P_i}{m_0-m_i}&\simeq -\frac{1}{|\Delta\gamma|^2}\diag(\Delta\gamma^*,\Delta\gamma^*)\mQ.
\end{align}
Then, the $i=2,3$ contributions are given by
\begin{align}
&-\frac{2}{|\Delta\gamma|^2}\Tr[P_0V\diag(\Delta\gamma^*,\Delta\gamma)\mQ V]\notag\\
&=-\frac{2}{|\Delta\gamma|^2}\rbraket{\psi_-|V\diag(\Delta\gamma^*,\Delta\gamma)\mQ V|\psi_-}\notag\\
&=-\frac{2}{|\Delta\gamma|^2}\Re[\Delta\gamma^*\tr[p_-\partial_q\halpha p_+\partial_q\halpha ]]\notag\\
&\simeq-\Re\left[\frac{2}{\Delta\lambda}\tr[p_-\partial_q\halpha p_+\partial_q\halpha ]\right].
\end{align}
We used $\Delta\gamma=\Delta\lambda+O(\epsilon)$.

Combining the terms obtained above, we arrive at
\begin{align}
\bar{m}_0''\simeq \partial_q^2\Re[\lambda_-]+\frac{\Re[\partial_q\lambda_-]^2}{\Re[\lambda_-]}+O(|\epsilon|^{1/2}).
\end{align}
Based on the criterion $m_0'' >0$, we conclude that the stability condition is generally given by $|\xi q|<1/\sqrt{3}$.

\subsection{Local stability with the current bias}
We have studied the fate of the small fluctuation around the steady state following the dynamics of the TDGL equation alone, that is, in the total absence of the electric field parallel to the system.
On the other hand, it is more experimentally feasible to consider the system with the current bias. 
This is achieved by imposing the current conservation~\cite{Artemenko1979-js}
\begin{align}
j_{\rm ex}
&=
2N_0\Re\left[\bm{\psi}^\dagger(x,t)v\bigl(\nabla/i\bigr)\bm{\psi}(x,t)\right]-\sigma_{\rm N}\nabla\bar{\phi}(x,t)
\notag\\
&=\text{const.},\label{eq:TDGL_current_fix}
\end{align}
with introducing the scalar potential $\bar{\phi}(x,t)$ into the TDGL equation.
Here, we adopt the gauge $\nabla\cdot\bar{{A}}(x,t)=0$ with $\lim_{x\to-\infty}\bar{A}(x,t)=0$, and thus the vector potential $\bar{A}(x,t)=0$ is omitted, while we adopt $\lim_{x\to-\infty}\bar{\phi}(x,t)=0$ for the scalar potential in the following.
For simplicity, we assume that $\halpha(\nabla/i)$ does not include $\nabla^{n}$ with $n\ge3$,
which allows us to use a simple expression for the supercurrent Eq.~\eqref{eq_SM:supercurrent} as given in the first term with defining $v(q)\equiv \partial_q\halpha(q)$.
The second term gives the contribution of the normal current induced by the electric field along the system and $\sigma_{\rm N}$ represents the normal conductivity. The TDGL equation
\begin{align}
&\frac{\Gamma}{N_0}[\partial_t+i\bar{\phi}(x,t)+i\Phi]\bm{\psi}(x,t)\notag\\
&=-[\halpha(\nabla/i)+\hat{\beta}(\bm{\psi}(x,t))]\bm{\psi}(x,t)
\end{align}
is to be solved simultaneously with Eq.~\eqref{eq:TDGL_current_fix}.

Let us consider the local stability of a steady state 
\begin{align}
&\bm{\psi}(x,t)=\bm{\psi}_\sst(q)e^{iqx+i\chi_\sst(q,t)},\quad \bar{\phi}(x,t)=0.
\end{align}
Here, $q$ is chosen to satisfy $j_{\rm ex}=j_\sst(q)$.
By considering the infinitesimal modification in the order parameter $\bm{\psi}(x,t)=e^{iqx+i\chi_\sst(q,t)}[\bm{\psi}_\sst(q)+\delta\bm{\psi}(x,t)]$ and also that in the
scalar potential
\begin{align}
\bar{\phi}(x,t)=\delta\bar{\phi}(x,t)=\int \frac{dk}{\sqrt{2\pi}}e^{ikx}\delta\phi_k(t),
\end{align}
the fixed-current condition gives
\begin{align}
\sigma\frac{\Gamma}{N_0}\nabla\delta\bar{\phi}(x,t)&=\Re\left[\bm{\psi}_\sst^\dagger v(q+\nabla/i)\delta\bm{\psi}(x,t)\right]\notag\\
&\quad+\Re\left[\delta\bm{\psi}^\dagger(x,t) v(q)\bm{\psi}_\sst(q)\right],
\end{align}
with $\sigma\equiv\sigma_N/2\Gamma$.
By using $\lim_{x\to-\infty}\bar{\phi}(x,t)=0$ and the Heaviside step function $\theta(x)=\int \frac{dk}{2\pi i}e^{ikx}(k-i\eta)^{-1}$ with $\eta=+0$, we obtain after integration
\begin{align}
\sigma\frac{\Gamma}{N_0}\delta\bar{\phi}(x,t)&=\int_{-\infty}^\infty dx'\,\theta(x-x')\sigma\frac{\Gamma}{N_0}\nabla'\delta\bar{\phi}(x',t)\notag\\
&=\int\frac{dk}{\sqrt{2\pi}}e^{ikx}(\bm{\psi}_\sst(q)^\dagger,\bm{\psi}_\sst^T(q))\label{eq:deltaphi_FT}\\
&\qquad\cdot\left[\frac{\tau_zV(q)}{ik+\eta}+\frac{\tau_z}{2i}W(q)\right]
\begin{pmatrix}
    \delta\bm{\psi}(k,t)\\
    \delta\bm{\psi}^*(-k,t)
\end{pmatrix},\notag
\end{align}
which gives $\delta\phi_k(t)$.
We used $v(q+k)=v(q)+kw(q)$, $w(q)\equiv\partial_q^2\halpha(q)$, $V(q)=\diag(v(q),-v(q)^*)$, and $W(q)=\diag(w(q),w(q)^*)$.
We assume that the perturbation of the order parameter is spatially localized.
Then, $\delta\bar{\phi}(x,t)$ given in Eq.~\eqref{eq:deltaphi_FT} generally converges to a finite value as $x\to\infty$, and $\delta\bar{\phi}(x,t)$ is always of the order of $\delta\bm{\psi}$.
Keeping first-order terms in the small quantities $\delta\bm{\psi}(x,t)$ and $\delta\bar{\phi}(x,t)$, and then performing the Fourier transform, the linearized TDGL equation is given by
\begin{align}
    \frac{\Gamma}{N_0}\partial_t\begin{pmatrix}
        \delta\bm{\psi}(k,t)\\
\delta\bm{\psi}^*(-k,t)
    \end{pmatrix}&=-M_k(q)\begin{pmatrix}
        \delta\bm{\psi}(k,t)\\
\delta\bm{\psi}^*(-k,t)
    \end{pmatrix}\notag\\
    &\qquad-i\frac{\Gamma}{N_0}\delta\phi_k(t)\begin{pmatrix}
        \bm{\psi}_\sst(q)\\
-\bm{\psi}^*_\sst(q)
    \end{pmatrix}\notag\\
    &=-L_k(q)\begin{pmatrix}
        \delta\bm{\psi}(k,t)\\
\delta\bm{\psi}^*(-k,t)
    \end{pmatrix},
\end{align}
by using $\delta\phi_{k}(t)=\delta\phi_{-k}^*(t)$ since $\bar{\phi}(x,t)$ is real.
In $L_k(q)\equiv M_k(q)+M_k^\phi(q)$, the first term is given by Eq.~\eqref{eq:Mkq_general}, while the second term additionally appears and is given by
\begin{align}
M_k^\phi(q)&\equiv 
\frac{2R_\sst(q)^2}{\sigma}\ket{\psi_-(q)}\bra{\psi_-(q)}\notag\\
&\qquad\quad\cdot\left[\frac{V(q)}{k-i\eta}+\frac{1}{2}W(q)\right].
\end{align}
Hereafter, the convergence factor $\eta$ is abbreviated since its role is just to specify the position where $\delta\bar{\phi}(x,t)$ vanishes.
We also keep the argument $q$ implicit for a while.

Our purpose is to know the condition to ensure that the real part of the eigenvalues of $L_k$ is positive for all $k\neq0$.
To obtain a more transparent expression, we perform a similarity transformation
$L_k\to B_kL_kB_k^{-1}$ with
\begin{align}
B_k&=Q_\psi+\sqrt{\frac{\sigma}{2R_\sst(q)^2}}k P_\psi,\notag\\
B_k^{-1}&=Q_\psi+\sqrt{\frac{2R_\sst(q)^2}{\sigma}}\frac{1}{k} P_\psi,
\end{align}
which does not change the eigenvalues.
Here, we defined
\begin{align}
P_\psi\equiv \ket{\psi_-}\bra{\psi_-},\quad Q_\psi=1-P_\psi,
\end{align}
which are generally different from $P_0=\ket{\psi_-}\rbra{\psi_-}$ due to the non-Hermitian nature of $\alphaeff$ (or $\hat{\gamma}$).  
This removes the infrared singularity of $M_k^\phi$ by
\begin{align}
B_kM_k^\phi B_k^{-1}&=\sqrt{\frac{2R^2_\sst}{\sigma}}P_\psi\left[V+\frac{k}{2}W\right]Q_\psi+\frac{R_\sst^2w_-}{\sigma}P_\psi\notag\\
&\to \sqrt{\frac{2R^2_\sst}{\sigma}}P_\psi VQ_\psi+\frac{R_\sst^2w_-}{\sigma}P_\psi,
\end{align}
as $k\to0$, where $w_-\equiv\braket{-|w|-}$.
We used the relation $\braket{\psi_-|V|\psi_-}=0$.
The other part of $L_k$, i.e., $M_k$, can be obtained by
\begin{align}
M_k=M_0+k\left[V+\frac{k}{2}W\right],
\end{align}
whose second term transforms by
\begin{align}
B_k\left(k\left[V+\frac{k}{2}W\right]\right)B_k^{-1}&\to \sqrt{\frac{2R_\sst^2}{\sigma}}Q_\psi VP_\psi,
\end{align}
for $k\to0$.
To see the behavior of the first term, note that
\begin{align}
M_0P_\psi=0,
\end{align}
since $\ket{\psi_-}$ is the zero mode of $M_0$.
This ensures that $B_kM_0B_k^{-1}$ is non-singular around $k=0$ and
\begin{align}
B_kM_0B_k^{-1}&\to Q_\psi M_0 Q_\psi. \quad(k\to 0)
\end{align}
Thus, the eigenvalues of $L_k$ in the limit $k\to0$ is given by those of
\begin{align}
\bar{L}_0&\equiv\lim_{k\to0}B_kL_kB_k^{-1}\notag\\
&=Q_\psi M_0Q_\psi+\frac{R_\sst^2w_-}{\sigma}P_\psi\notag\\
&\quad+\sqrt{\frac{2R_\sst^2}{\sigma}}[Q_\psi VP_\psi+P_\psi VQ_\psi].
\end{align}
By using the spectral decomposition of $\mM_0$ given in Eq.~\eqref{eq:SD_M0}, that of $Q_\psi M_0Q_\psi$ is given by
\begin{align}
\!\!\! Q_\psi M_0 Q_\psi=0\cdot P_0'+2\beta_-R_\sst^2P_1'+\Delta\gamma P_2'+\Delta\gamma^*P_3',
\end{align}
neglecting $O(\epsilon^2)$ terms,
where it is easy to see that $P_0'\equiv P_\psi$ and $P_i'\equiv Q_\psi e^{-S}P_ie^{S}$ $(i=1,2,3)$ satisfy $P_a'P_b'=\delta_{ab}P_a'$ for $a,b=0,1,2,3$ and $\sum_aP_a'=1$.
Thus, we obtain
\begin{align}
\bar{L}_0&=\frac{R_\sst^2w_-}{\sigma}P_0'+
2\beta_-R^2_\sst P_1'+\Delta\gamma P_2'+\Delta\gamma^* P_3'\notag\\
&\qquad+\sqrt{\frac{2R_\sst^2}{\sigma}}\sum_{i\neq0}[P_i'VP_0'+P_0'VP_i'].
\end{align}

By writing $P_i'=\ket{i'}\rbra{i'}$, the matrix $\bar{L}_0$ is represented by $\bar{L}_0=\ket{i'}[\bar{l}_0]_{ij}\rbra{j'}$ with
\begin{widetext}
    \begin{align}
\bar{l}_0&=\begin{pmatrix}
    \frac{R_\sst^2w_-}{\sigma}&\sqrt{\frac{2R_\sst^2}{\sigma}}\rbraket{0'|V|1'}&\sqrt{\frac{2R_\sst^2}{\sigma}}\rbraket{0'|V|2'}&\sqrt{\frac{2R_\sst^2}{\sigma}}\rbraket{0'|V|3'}\\
    \sqrt{\frac{2R_\sst^2}{\sigma}}\rbraket{1'|V|0'}&2\beta_-R_\sst^2&0&0\\
    \sqrt{\frac{2R_\sst^2}{\sigma}}\rbraket{2'|V|0'}&0&\Delta\gamma&0\\
    \sqrt{\frac{2R_\sst^2}{\sigma}}\rbraket{3'|V|0'}&0&0&\Delta\gamma^*
\end{pmatrix}.
\end{align}
\end{widetext}
We obtain after
another similarity transformation by $B'=\diag\left(1,\frac{v_{10}}{[\bar{l}_0]_{10}},\frac{v_{20}}{[\bar{l}_0]_{20}},\frac{v_{30}}{[\bar{l}_0]_{30}}\right)$,
\begin{align}
\!\!\! B'\bar{l}_0(B')^{-1}&= l_0\equiv\begin{pmatrix}
    \frac{R_\sst^2w_-}{\sigma}&v_{10}&v_{20}&v_{30}\\
    v_{10}&2\beta_-R_\sst^2&0&0\\
    v_{20}&0&\Delta\gamma&0\\
   v_{30}&0&0&\Delta\gamma^*
\end{pmatrix},
\end{align}
where $v_{i0}$ is a root of 
\begin{align}
v_{i0}^2&=\bar{l}_{0i}\bar{l}_{i0}\\
&=\frac{2R_\sst^2}{\sigma}\rbraket{0'|V|i'}\rbraket{i'|V|0'}\notag\\
&=\frac{2R_\sst^2}{\sigma}\Tr[P_0'VP_i'V].
\end{align}
Specifically, $v_{10}^2$ is given by
\begin{align}
v_{10}^2&=\frac{2R_\sst^2}{\sigma}\Tr[P_\psi VQ_\psi e^{-S}P_1e^S V]\notag\\
&=\frac{2R_\sst^2}{\sigma}\Bigl(\Tr[P_\psi VP_1V]+O(|\epsilon|^{3/2})\Bigr)\notag\\
&=\frac{2R_\sst^2}{\sigma}\Re[\braket{-|v|-}]\Re[\rbraket{-|v|-}]+O(|\epsilon|^{5/2}),
\end{align}
while $v_{20}^2$ and $v_{30}^2$ are given by
\begin{align}
v_{20}^2&=\frac{2R^2_\sst}{\sigma}\Tr[P_\psi Ve^{-S}P_2e^SV]\notag\\
&=\frac{2R^2_\sst}{\sigma}\Tr[P_\psi VP_2V]+O(\epsilon^2)\notag\\
&=\frac{R_\sst^2}{\sigma}\braket{-|v|+}\rbraket{+|v|-}+O(\epsilon^2),\\
v_{30}^2&=\frac{R_\sst^2}{\sigma}[\braket{-|v|+}\rbraket{+|v|-}]^*+O(\epsilon^2).
\end{align}
Thus, we obtain $v_{10}=O(\epsilon)$ while $v_{20},v_{30}=O(|\epsilon|^{1/2})$.

Now, the problem of the long-wavelength stability recasts into the eigenproblem of $l_0$.
Among the matrix elements of $l_0$, $\Delta\gamma$ and $\Delta\gamma^*$ have the largest magnitude of $O(1)$ while the others vanish as $\epsilon\to0$.
We are interested in the eigenvalue with the smallest real part, which would be $O(\epsilon)$.
Thus, we first diagonalize $l_0$ with neglecting $v_{20}$ and $v_{30}$ and then take them into account.
Since $v_{20}$ and $v_{30}$ are $O(|\epsilon|^{1/2})$, we adopt the (nearly-)degenerate second-order perturbation theory (or equivalently the Schiffer-Wolff transformation). The eigenvalues of $O(\epsilon)$ are given by diagonalizing the effective matrix
\begin{align}
l_{\rm eff}\equiv&\begin{pmatrix}
    \frac{R_\sst^2w_-}{\sigma}&v_{10}\\
    v_{10}&2\beta_-R_\sst^2
\end{pmatrix}-\begin{pmatrix}
   v_{20}&v_{30}\\
0&0
\end{pmatrix}\notag\\
&\qquad\qquad\cdot
\begin{pmatrix}
    \frac{1}{\Delta\gamma}&0\\
    0&\frac{1}{\Delta\gamma^*}
\end{pmatrix}
\begin{pmatrix}
   v_{20}&0\\
v_{30}&0
\end{pmatrix}+O(\epsilon^2)\notag\\
&=\begin{pmatrix}
    \frac{R_\sst^2\partial_q^2\Re\lambda_-}{\sigma}&v_{10}\\
    v_{10}&2\beta_-R_\sst^2
\end{pmatrix}+O(\epsilon^{3/2}),
\end{align}
where we used
\begin{align}
{w_-}&{=\partial_q^2\Re\lambda_-+\frac{v_{20}^2}{\Delta\gamma}+\frac{v_{30}^2}{\Delta\gamma^*}+O(|\epsilon|^{1/2}),}
\end{align}
since the difference of $\bra{-}$ and $\rbra{-}$ is $O(|\epsilon|^{1/2})$ and that of the eigenstates of $\hat{\gamma}$ and $\alphaeff$ is $O(\epsilon)$.
Finally, we obtain the eigenvalue with the lowest real part
\begin{align}
\lambda_{0}&=\frac{R_\sst^2\partial_q^2\Re\lambda_-}{2\sigma}+\beta_-R_\sst^2\notag\\
&\quad-\sqrt{\left(\frac{R_\sst^2\partial_q^2\Re\lambda_-}{2\sigma}-\beta_-R_\sst^2\right)^2+v_{10}^2}\notag\\
&=\beta_-R^2_\sst\left[1+\frac{u}{2}-\sqrt{\left(1-\frac{u}{2}\right)^2+2us}\right],
\end{align}
with $u\equiv\frac{\partial_q^2\Re\lambda_-}{\sigma\beta_-}>0$ and
\begin{align}
s&\equiv\frac{\braket{-|v|-}\Re[\rbraket{-|v|-}]}{\beta_-^2R_\sst^2\partial_q^2\Re\lambda_-}\notag\\
&=\frac{\braket{-|v|-}\partial_q\Re\lambda_-}{-\Re\lambda_-\,\partial_q^2\Re\lambda_-}.
\end{align}
The sign of $\Re\lambda_0=\lambda_0$ is determined by that of
\begin{align}
(1+u/2)^2-(1-u/2)^2-2us&=2u(1-s),
\end{align}
and thus we conclude
\begin{align}
\sgn\Re[\lambda_0]=\sgn[1-s].
\end{align}
For our model, we have
\begin{subequations}
    \begin{align}
&\Re\lambda_-/|\bar{\epsilon}(\mE)|=-1+(\xi q)^2,\\
&\braket{-|v|-}/|\bar{\epsilon}(\mE)|=2\xi^2(q-q_\mE).
\end{align}
\end{subequations}
Thus, we obtain
\begin{align}
&[-\Re\lambda_-\partial_q^2\Re\lambda_-](1-s)\notag\\
&=-\Re[\lambda_-]\partial_q\braket{-|v|-}-\braket{-|v|-}\partial_q\Re[\lambda_-]\notag\\
&=\partial_q\Bigl[-\Re[\lambda_-]\braket{-|v|-}\Bigr],
\end{align}
and conclude
\begin{align}
\sgn[1-s]=\sgn\frac{dj_\sst(q)}{dq}.
\end{align}
The steady state is locally stable only when $dj_\sst(q)/dq>0$.

We have shown that the eigenvalues of $L_k$ are positive for $k\to0$ as long as $s<1$.
Strictly speaking, $dj_\sst(q)/dq>0$ is a necessary condition for the local stability and we have to check that the eigenvalues do not change sign for $k>0$.
The $k$ dependence of $\lambda_0$ is given by
\begin{align}
\frac{\lambda_0(k)}{\beta_-R_\sst^2}&=1+\frac{u}{2}+u\bk^2-ic'\bk\notag\\
&\quad-\sqrt{(1-u/2-ic'\bk)^2+2us(1+\bk^2)},\label{eq:eigenvalue_kdep}
\end{align}
by repeating the above discussion with keeping $k$, whose details are given in the next subsection.
Here, we define the scaled wave number $\bk$ and a constant $c'$ by
\begin{align}
k=\sqrt{\frac{2R_\sst^2}{\sigma}}\bk,\quad
c'=\frac{1}{2}\sqrt{\frac{\sigma}{2R_\sst^2}}\Im[\braket{-|+}\rbraket{+|v|-}],
\end{align}
where $c'$ is $O(1)$ in terms of the reduced temperature $\epsilon$.
After cumbersome calculations detailed in the next subsection, we can show that $\Re[\lambda_0(k)]$ is positive for all $k$ as long as $s<1$.
Thus, $s<1$, i.e., $dj_\sst(q)/dq>0$ gives the local stability condition of the steady state under the current bias.

Precisely speaking, the stability condition obtained in this section is for a superconducting wire whose width is smaller than the coherence length since a one-dimensional system is assumed while the transverse coordinate and the magnetic field are not taken into account.
The stability analysis discussed in this section would be generalized to thin films whose width is smaller than the Pearl length $\lambda^2/d$ with the thickness $d$ and the penetration depth $\lambda$, where the magnetic field produced by the electric current can be neglected and the spatially uniform configuration of the electric current gives a good approximation.

\subsection{Details of the derivation of Eq.~\eqref{eq:eigenvalue_kdep}}
Here we show the details of the derivation of Eq.~\eqref{eq:eigenvalue_kdep} and the proof of $\Re[\lambda_0(k)]>0$.
Following the procedure of the previous section without taking the limit $k\to0$, we obtain
    \begin{align}
    &B_kL_kB_k^{-1}\notag\\
    &=\frac{R^2_\sst w_-(1+\bk^2)}{\sigma}P_\psi\notag\\
    &\quad+Q_\psi\left[M_0+\sqrt{\frac{2R^2_\sst}{\sigma}}\bk V+{\frac{R^2_\sst}{\sigma}}\,\bk^2 W\right]Q_\psi\notag\\
    &\quad+\sqrt{\frac{2R^2_\sst}{\sigma}}(1+\bk^2)P_\psi\left[V+\sqrt{\frac{2R^2_\sst}{\sigma}}\frac{\bk}{2}W\right]Q_\psi\notag\\
    &\quad+\sqrt{\frac{2R_\sst^2}{\sigma}}Q_\psi\left[V+\sqrt{\frac{2R^2_\sst}{\sigma}}\frac{\bk}{2}W\right]P_\psi\notag\\
    &\quad+\bk P_\psi  M_0 Q_\psi.
    \end{align}
Another similarity transformation $B'_k\equiv \frac{1}{\sqrt{1+\bk^2}}P_\psi+Q_\psi$ yields
    \begin{align}
    &B_k'B_kL_kB_k^{-1}(B_k')^{-1}\notag\\
    &=\frac{R^2_\sst w_-(1+\bk^2)}{\sigma}P_\psi\notag\\
    &\quad+Q_\psi\left[M_0+\sqrt{\frac{2R^2_\sst}{\sigma}}\bk V+{\frac{R^2_\sst}{\sigma}}\,\bk^2 W\right]Q_\psi\notag\\
    &\quad+\sqrt{\frac{2R^2_\sst}{\sigma}}\sqrt{1+\bk^2}P_\psi\left[V+\sqrt{\frac{2R^2_\sst}{\sigma}}\frac{\bk}{2}W\right]Q_\psi\notag\\
    &\quad+\sqrt{\frac{2R_\sst^2}{\sigma}}\sqrt{1+\bk^2}Q_\psi\left[V+\sqrt{\frac{2R^2_\sst}{\sigma}}\frac{\bk}{2}W\right]P_\psi\notag\\
    &\quad+\frac{\bk}{\sqrt{1+\bk^2}} P_\psi  M_0 Q_\psi.
    \end{align}
We decompose this in terms of the order of $\epsilon$,
\begin{align}
&B_k'B_kL_kB_k^{-1}(B_k')^{-1}\notag\\
&=L_{k,0}+L_{k,1/2}+L_{k,1}+O(|\epsilon|^{3/2}),
\end{align}
where $L_{k,i}$ is $O(|\epsilon|^i)$ and $\bk$ is regarded as $O(1)$.
The matrix $L_{0,k}$ is given by
\begin{align}
L_{0,k}&=\Delta\gamma P_2'+\Delta\gamma^*P_3'\sim
\begin{pmatrix}
    0&0&0&0\\
    0&0&0&0\\
    0&0&\Delta\gamma&0\\
    0&0&0&\Delta\gamma^*
\end{pmatrix},
\end{align}
where ``$\sim$" means the representation in the basis $\ket{i'}\rbra{j'}$ (this is also a similarity transformation).
By noting that $c=O(|\epsilon|^{1/2})$, we obtain $L_{k,1/2}$ by
\begin{align}
L_{k,1/2}&=\sqrt{\frac{2R^2_\sst}{\sigma}}\bk\,[P_1' V(P_2'+P_3')+(P_2'+P_3')VP_1']\notag\\
&\quad+\frac{\bk}{\sqrt{1+\bk^2}} P_0'  M_0 (P_2'+P_3')\notag\\
&\quad+\sqrt{\frac{2R^2_\sst}{\sigma}}\sqrt{1+\bk^2}\Bigl[P_0' V(P_2'+P_3')\notag\\
&\qquad+(P_2'+P_3')VP_0'\Bigr].
\end{align}
This is of the form
\begin{align}
L_{k,1/2}\sim
\begin{pmatrix}
\begin{matrix}0&0\\
0&0
\end{matrix}&\large{\ l_{PQ}\ }\\
\large{\ l_{QP}\ }&\begin{matrix}0&0\\
0&0
\end{matrix}
\end{pmatrix}
\equiv\begin{pmatrix}
0&0&v'_{02}&v'_{03}\\
    0&0&v'_{12}&v'_{13}\\
    v'_{20}&v'_{21}&0&0\\
   v'_{30}&v'_{31}&0&0
\end{pmatrix}.
\end{align}
The matrix $L_{k,1}$ is given by
\begin{align}
L_{k,1}&=\frac{R^2_\sst w_-(1+\bk^2)}{\sigma}P_0'+2\beta_-R_\sst^2P_1'\notag\\
&\quad
+\sqrt{\frac{2R^2_\sst}{\sigma}}\bk\,[P_1' VP_1']
+\frac{R_\sst^2\bk^2}{\sigma}P_1'WP_1'\notag\\
&\quad+\sqrt{\frac{2R_\sst^2}{\sigma}}\sqrt{1+\bk^2}\left[P_0'\left(V+\sqrt{\frac{2R_\sst^2}{\sigma}}\frac{\bk}{2}W\right)P_1'\right.\notag\\
&\qquad\left.+P_1'\left(V+\sqrt{\frac{2R_\sst^2}{\sigma}}\frac{\bk}{2}W\right)P_0'\right]\notag\\
&\quad+{\frac{\bk}{\sqrt{1+\bk^2}}P_0'M_0P_1'}+[*]
\notag\\
&=\frac{R^2_\sst w_-(1+\bk^2)}{\sigma}P_0'+2\beta_-R_\sst^2P_1'
+\frac{R_\sst^2\bk^2}{\sigma}P_1'WP_1'\notag\\
&\quad+\sqrt{\frac{2R_\sst^2}{\sigma}}\sqrt{1+\bk^2}\left[P_0'VP_1'+P_1'VP_0'\right]+[*],
\end{align}
where ``$*$" indicates the matrix elements irrelevant to the results.
Here, we used the relations
\begin{align}
\!\!\! P_1'VP_1'=O(\epsilon),\  
P_0'WP_1'=O(\epsilon), \ 
P_0'M_0P_1'=O(\epsilon),
\end{align}
by noting that we can use $P_i$ instead of $P_i'$ for the evaluation of $L_{k,1}$.
The matrix $L_{k,1}$ is of the form
\begin{align}
\!\!\!\!L_{k,1}&\sim\begin{pmatrix}
    \frac{R_\sst^2w_-(1+\bk^2)}{\sigma}&v_{01}'&*&*\\
    v_{10}'&2\beta_-R_\sst^2\left(1+\frac{\bk^2w_-}{2\beta_-\sigma}\right)&*&*\\
    *&*&*&*\\
   *&*&*&*
\end{pmatrix},
\end{align}
by using
\begin{align}
P_1'WP_1'=w_-P_1'+O(\epsilon).
\end{align}

The effective matrix in the small-eigenvalue subspace is given by
\begin{align}
l_{\rm eff}(k)&=\begin{pmatrix}
    \frac{R_\sst^2w_-(1+\bk^2)}{\sigma}&v'_{01}\\
    v'_{10}&2\beta_-R_\sst^2\left(1+\frac{\bk^2w_-}{2\beta_-\sigma}\right)
\end{pmatrix}\notag\\
&\quad-l_{PQ}\diag(1/\Delta\gamma,1/\Delta\gamma^*)l_{QP}\notag\\
&=\begin{pmatrix}
    \frac{R_\sst^2w_-(1+\bk^2)}{\sigma}&v'_{01}\\
    v'_{10}&2\beta_-R_\sst^2\left(1+\frac{\bk^2w_-}{2\beta_-\sigma}\right)
\end{pmatrix}\notag\\
&\qquad-\begin{pmatrix}
\frac{v'_{02}v'_{20}}{\Delta\gamma}+\frac{v'_{03}v'_{30}}{\Delta\gamma^*}&\frac{v'_{02}v'_{21}}{\Delta\gamma}+\frac{v'_{03}v'_{31}}{\Delta\gamma^*}\\
\frac{v'_{12}v'_{20}}{\Delta\gamma}+\frac{v'_{13}v'_{30}}{\Delta\gamma^*}&\frac{v'_{12}v'_{21}}{\Delta\gamma}+\frac{v'_{13}v'_{31}}{\Delta\gamma^*}    
\end{pmatrix}.
\end{align}

Let us calculate the matrix elements $v_{ij}'$.
We have
\begin{subequations}
\begin{align}
\rbraket{0'|M_0|2'}&\simeq \braket{0|M_0|2}\notag\\
&=\frac{1}{\sqrt{2}}(\bra{-},-\bra{-}^*)\begin{pmatrix}
    \Delta\gamma p_+&0\\
    0&\Delta\gamma^*p_+^*
\end{pmatrix}\begin{pmatrix}
    \ket{+}\\
    0
\end{pmatrix}\notag\\
&=\frac{\Delta\gamma}{\sqrt{2}}c,\\
\rbraket{0'|M_0|3'}&=-\frac{\Delta\gamma^*}{\sqrt{2}}c,
\end{align}\end{subequations}
where $c\equiv\braket{-|+}$ is chosen to be real,
and thus we obtain by using
$v_{+-}\equiv\braket{+|v|-}\simeq\rbraket{+|v|-}$,
\begin{align}
&\frac{v'_{02}v'_{20}}{\Delta\gamma}+\frac{v'_{03}v'_{30}}{\Delta\gamma^*}\\
&=\frac{1}{\Delta\gamma}\left(\sqrt{\frac{2R_\sst^2(1+\bk^2)}{\sigma}}\rbraket{0'|V|2'}\right.\notag\\
&\left.\qquad\qquad\qquad+\frac{\bk}{\sqrt{1+\bk^2}}\frac{\Delta\gamma c}{\sqrt{2}}\right)\sqrt{\frac{2R_\sst^2(1+\bk^2)}{\sigma}}\rbraket{2'|V|0'}\notag\\
&\quad+\frac{1}{\Delta\gamma^*}\left(\sqrt{\frac{2R_\sst^2(1+\bk^2)}{\sigma}}\rbraket{0'|V|3'}\right.\notag\\
&\qquad\qquad\qquad\left.-\frac{\bk}{\sqrt{1+\bk^2}}\frac{\Delta\gamma^* c}{\sqrt{2}}\right)\sqrt{\frac{2R_\sst^2(1+\bk^2)}{\sigma}}\rbraket{3'|V|0'}\notag\\
&=|v_{+-}|^2\frac{R_\sst^2(1+\bk^2)}{\sigma}\left(\frac{1}{\Delta\gamma}+\frac{1}{\Delta\gamma^*}\right)+i\bk\Im[cv_{+-}]\sqrt{\frac{2R_\sst^2}{\sigma}}.\notag
\end{align}
We also obtain
\begin{align}
\frac{v'_{12}v'_{21}}{\Delta\gamma}+\frac{v'_{13}v'_{31}}{\Delta\gamma^*}
&=|v_{+-}|^2\frac{R_\sst^2\bk^2}{\sigma}\left(\frac{1}{\Delta\gamma}+\frac{1}{\Delta\gamma^*}\right),
\end{align}
and define
\begin{subequations}\begin{align}
\beta_-R_\sst^2 \bar{v}_{01}&=v'_{01}+\frac{v'_{02}v'_{21}}{\Delta\gamma}+\frac{v'_{03}v'_{31}}{\Delta\gamma^*}\notag\\
&\simeq\sqrt{\frac{2R_\sst^2(1+\bk^2)}{\sigma}}\braket{-|v|-},\\
\beta_-R_\sst^2 \bar{v}_{10}&=v'_{10}+\frac{v'_{12}v'_{20}}{\Delta\gamma}+\frac{v'_{13}v'_{30}}{\Delta\gamma^*}\notag\\
&\simeq\sqrt{\frac{2R_\sst^2(1+\bk^2)}{\sigma}}\rbraket{-|v|-},
\end{align}\end{subequations}
by using $\Im[\Delta\gamma]=O(|\epsilon|^{1/2})$.
Thus, we obtain
\begin{widetext}
\begin{align}
l_{\rm eff}(k)&=\begin{pmatrix}
    \frac{R_\sst^2\partial_q^2\lambda_-}{\sigma}(1+\bk^2)-i\bk\Im[cv_{+-}]\sqrt{\frac{2R_\sst^2}{\sigma}}&\beta_-R_\sst^2 v_{01}'\\
    \beta_-R_\sst^2 v_{10}'&2\beta_-R_\sst^2\left(1+\frac{\bk^2\partial_q^2\lambda_-}{2\beta_-\sigma}\right)
\end{pmatrix}
=\beta_-R_\sst^2 \begin{pmatrix}
    u(1+\bk^2)-i2c'\bk&v_{01}'\\
    v_{10}'&2+u\bk^2
\end{pmatrix},
\end{align}
\end{widetext}
with
\begin{align}
2c'=\sqrt{\frac{\sigma}{2R_\sst^2}}\Im[cv_{+-}]=O(1).
\end{align}
The eigenvalue with the lowest real part is given by
\begin{align}
\frac{\lambda_0(k)}{\beta_-R_\sst^2}&=1+\frac{u}{2}+u\bk^2-ic'\bk\notag\\
&\quad-\sqrt{(1-u/2-ic'\bk)^2+2us(1+\bk^2)}.
\end{align}

In the following, we show that $\Re[\lambda_0(k)]>0$ when $s<1$ is satisfied.
We can say $\Re[\lambda_0(k)]>0$ if 
\begin{align}
&\left(1+\frac{u}{2}+u\bk^2\right)^2\\
&\qquad-\Re\left[\sqrt{(1-\frac{u}{2}-ic'\bk)^2+2us(1+\bk^2)}\right]^2\notag\\
&=\left[\left(1+\frac{u}{2}+u\bk^2\right)^2-\frac{(1-\frac{u}{2})^2-(c'\bk)^2+2us(1+\bk^2)}{2}\right]\notag\\
&\quad-\frac{\sqrt{[(1-\frac{u}{2})^2-(c'\bk)^2+2us(1+\bk^2)]^2+[2c'\bk(1-\frac{u}{2})]^2}}{2},\notag
\end{align}
is positive.
The first term is positive for $s<2$ since
\begin{align}
&\left(1+\frac{u}{2}+u\bk^2\right)^2-\frac{(1-\frac{u}{2})^2-(c'\bk)^2+2us(1+\bk^2)}{2}\\
&=(1+\bk^2)(u\bk^2+(2-s)u)+\frac{1}{2}\left(1-\frac{u}{2}\right)^2+(c'\bk)^2>0.\notag
\end{align}
Thus, $\Re[\lambda_0(k)]>0$ follows for $s<1$ since 
\begin{align}
&\left[\left(1+\frac{u}{2}+u\bk^2\right)^2-\frac{(1-u/2)^2-(c'\bk)^2+2us(1+\bk^2)}{2}\right]^2\notag\\
&\qquad-\frac{[(1-u/2)^2-(c'\bk)^2+2us(1+\bk^2)]^2+[2c'\bk(1-u/2)]^2}{4}\notag\\
&=(1+\bk^2)(u^2\bk^2+2u(1-s))\left(1+\frac{u}{2}+u\bk^2\right)^2\notag\\
&\qquad\qquad+(1+\bk^2)(u^2\bk^2+2u)(c'\bk)^2>0.
\end{align}

\end{document}